\documentclass[
 reprint,
 amsmath,amssymb,
 aps,unsortedaddress]{revtex4-1}
 \usepackage{xcolor}
\usepackage{graphicx}% Include figure files
\usepackage{dcolumn}% Align table columns on decimal point
\usepackage{bm}% bold math

\newcommand{\be}{\begin{equation}}
\newcommand{\ee}{\end{equation}}
\newcommand{\n}{\nonumber \\ }
\begin{document}
\graphicspath{{figures/}}

\title{Subsystem symmetry protected topological order}

\author{Yizhi You}
\affiliation{Princeton Center for Theoretical Science, Princeton University, 
NJ, 08544, USA}

\author{Trithep Devakul}
\affiliation{Department of Physics, Princeton University, 
NJ, 08544, USA}

\author{F.~J. Burnell}
\affiliation{Department of Physics, University of Minnesota Twin Cities, 
MN, 55455, USA}

\author{S.~L. Sondhi}
\affiliation{Department of Physics, Princeton University, 
NJ 08544, USA}

\date{\today}
\begin{abstract}
In this work, we introduce a new type of topological order which is protected by subsystem symmetries which act on lower dimensional subsets of lattice many-body system, e.g. along lines or planes in a three dimensional system. The symmetry groups for such systems exhibit a macroscopic number of generators in the infinite volume limit. We construct a set of exactly solvable models in $2D$ and $3D$ which exhibit such subsystem SPT (SSPT) phases with one dimensional subsystem symmetries. These phases exhibit analogs of phenomena seen in SPTs protected by global symmetries: gapless edge modes, projective realizations of the symmetries at the edge and non-local order parameters. Such SSPT phases are proximate, in theory space, to previously studied phases that break the subsystem symmetries and phases with fracton order which result upon gauging them. 

\end{abstract}

\maketitle

\section{Introduction}

Symmetry plays a pivotal role in distinguishing phases of matter.  The great majority of the phases seen in nature are distinguished by different patterns of spontaneous symmetry breaking. Recently, it has been appreciated that multiple phases with the same {\it unbroken} global symmetry can also exist.  The new class of phases, which realize the unbroken global symmetry in distinct ways, are known as symmetry-protected topological (SPT) phases of matter.  

The existence of multiple phases with the same unbroken symmetry was first recognized for integer spin chains~\cite{Haldane1983-ya,Affleck1987-wn} and polyacetylene~\cite{Su1979-rl}, and generalized to any symmetry group in 1D~\cite{Schuch2011-jx,Turner2011-zi,fidkowski2011topological,Pollmann2012-lv,Chen2011-et}; such phases are characterized by symmetry protected gapless boundary modes under open boundary conditions.
A similar phenomenon occurs in higher dimensions in band insulators~\cite{Kane2005-ml,Moore2007-jk,Fu2007-xo}, interacting systems of bosons~\cite{Chen2011-kz,Chen2012-oa,Levin2012-dv,Vishwanath2013-pb} and fermions~\cite{Yao2013-vc,Gu2014-lj,Qi2013-bl,Cheng2015-ul,Gaiotto2016-ba,thorngren2016gauging,else2014classifying}.
The unifying features of such phases are unusual boundary modes whose existence is guaranteed as long as the symmetry is unbroken.

A second class of phases beyond the broken symmetry paradigm are the ``topologically ordered''\cite{wen1990topological} phases like superconductors\cite{hansson2004superconductors}, spin liquids and quantum Hall phases which exhibit fractionalization in the bulk. These exhibit an unbroken local symmetry/gauge invariance which is either present microscopically or is emergent in the region of parameter space where they are absolutely stable\cite{von2016absolute}.

However, this seemingly comprehensive picture of how unbroken symmetry---global or local---can lead to distinct phases of matter misses an interesting intermediate possibility, known as subsystem symmetry.  A subsystem symmetry consists of independent symmetry operations acting on an extensively large set of $d$-dimensional subsystems, with $0<d<D$ for a fixed $D$. For example, a $d=2$ subsystem symmetry acts on a planar region in the $D=3$ dimensional system, while a $d=1$ symmetry acts along a line.
As such, they have also referred to as \emph{intermediate} or \emph{gauge-like} symmetries, as they interpolate between global ($d=D$) symmetries and local ($d=0$) gauge symmetries. 
Theories with such symmetries may display dimensional reduction~\cite{Batista2005-yr} and arise, for example, in models of spin and orbital degrees of freedom, such as the Kugel-Khomskii model~\cite{Kugel1973-bf,Van_den_Brink2004-og}; from Jahn-Teller effects~\cite{Van_den_Brink2004-og}; and in orbital compass models~\cite{Xu2004-oj} ---
the last of which in two dimensions is dual to the Xu-Moore model of $p\pm ip$ superconducting arrays~\cite{Xu2004-oj,Xu2005-df} (which we will introduce in more detail later).

Subsystem symmetries have recently become a subject of interest from an orthogonal direction, when it was discovered that, in 3D, applying a generalized gauging procedure~\cite{Williamson2016-lv,Vijay2016-dr} to models with such symmetries resulted in theories with \emph{fracton} order~\cite{Vijay2016-dr,Chamon2005-fc,Haah2011-ny,Bravyi2011-fl,Yoshida2013-of,Vijay2015-jj} --- novel phases of matter characterized by subextensive topological ground state degeneracy and quasiparticle excitations with restricted mobility which have been the subject of much recent research~\cite{Slagle2017-la, Petrova2017-pe, Slagle2017-gk, Devakul2017-gg, Ma2017-cb, He2017-eq, Schmitz2017-ky, Slagle2017-st, Vijay2017-ey, Halasz2017-ov, Slagle2017-ne, Shi2017-bs, Ma2017-qq, Vijay2016-dr, Hsieh2017-sc,Pretko2017-ej,Pretko2017-nt,Prem2017-ql,pretko2016entanglement}.

What are the possible phases of a model exhibiting subsystem symmetry?  
It is well understood through Elitzur's theorem~\cite{Elitzur1975-wh} that zero-dimensional (local or \emph{gauge}) symmetries cannot be spontaneously broken and gauge non-invariant observables have strictly zero expectation value.  As noted above such $d=0$ symmetries can, however, lead to topologically ordered phases, which are stable to arbitrary small perturbations if the spectrum is gapped \cite{bravyi2010topological}.
For $d>0$ dimensional symmetries, symmetry breaking is possible and a generalized Elitzur's theorem~\cite{Batista2005-yr} instead bounds the expectation value of symmetry non-invariant observables by those of a $d$-dimensional model.  
Thus, $d>0$ dimensional discrete symmetries can be spontaneously broken and a concrete example, where a $d=2$ subsystem Ising symmetry can be spontaneously broken  at low temperatures, was given in Ref.~\cite{Johnston2016-nf}.

In this paper we ask whether systems with subsystem symmetry likewise admit multiple distinct symmetric phases in which the symmetry is not spontaneously broken --- which we call subsystem symmetry protected topological (SSPT) phases and find that the answer is in the affirmative.
Specifically, we focus on $d=1$ subsystem symmetries; in a companion paper we will treat the case of $d=2$. For these we construct models in three distinct classes: (i) for bosons with unitary subsystem symmetries, (ii) for bosons with subsystem symmetries and a non-unitary time-reversal symmetry and,
(iii) for fermions with subsystem fermion parity conservation and a global time-reversal symmetry. The SSPT phases in these models exhibit various interesting properties including entangled ground-states, protected gapless boundary modes, and a non-local order parameter. These properties are closely analogous to those of SPT phases, where the unbroken symmetry is global~\cite{Gu2009-ly,chen2014symmetry,chen2011two,fidkowski2010effects,metlitski2013bosonic}. We also demonstrate that our phases are distinct from ``weak'' SSPT phases constructed by suitably stacking $1D$ SPT chains each with their own global symmetry, and weakly coupling them in a manner respecting the subsystem symmetry. Finally for systems in class (i) we analyze a particular perturbation that takes us out of the SSPT phase via a duality transformation.

The paper is organized as follows. In Section~\ref{sec:2dplaq}, we introduce a topological plaquette paramagnet, previously discussed in the context of cluster states\cite{Raussendorf2001-xm,Doherty2009-fd}, which has gapless non-dispersing boundary modes protected by $1d$ $Z^{sub}_2$ subsystem symmetry.   
We identify a non-local membrane-like bulk order parameter that detects the ``decorated defect condensate"\cite{you2016stripe,you2016decorated,chen2014symmetry} nature of the ground state, and thus distinguishes the topological and trivial phases. In Section \ref{3dversion}, we show that a similar situation exists for higher dimensional SSPTs with $1d$ subsystem symmetries. Specifically, we introduce a model with a 3D SSPT phase protected by $1d$ $Z^{sub}_2 $ symmetry with protected gapless surface modes and a non-zero volume-like order parameter. 
Next we construct two types of exactly solvable models with SSPT order protected by an anti-unitary symmetry. In Section \ref{timesub}, we introduce a spin system with subsystem time-reversal symmetry $\mathcal{T}^{sub}$ in both $2D$ and $3D$. Akin to the valence-bond ground state of the $\mathcal{T}$ invariant AKLT chain, the ground state of this SSPT can be regarded as a valence plaquette solid($2D$) or valence cube solid($3D$) with maximal entanglement in each plaquette ($2D$) or cube ($3D$). Finally, in Section~\ref{fermionsub}, we turn to fermionic systems, constructing an exactly solvable model in 2$D$ with subsystem fermion parity symmetry and time reversal via the Fidkowski-Kitaev  interaction\cite{fidkowski2010effects}. We show that in this model the combined fermion parity and time reversal symmetries guarantee the existence of a gapless, non-dispersing boundary mode.

\section{$Z_2^{sub}$ $1d$ symmetry in $2D$}\label{sec:2dplaq}

\subsection{Trivial paramagnet}

We start by reviewing the Xu-Moore model~\cite{Xu2004-oj,Xu2005-df}, which we will refer to as the ``Plaquette Ising model'' (PIM).  The model consists of Ising spins on the sites of an $L\times L$ square lattice, governed by the Hamiltonian,
\begin{equation} 
H_\text{PIM}=-\sum_{ijkl \in P} \sigma^z_i \sigma^z_j \sigma^z_k \sigma^z_l -\Gamma \sum_{i}\sigma_i^x
\label{pising}
\end{equation}
where $\sigma^{z,x}_i$ are Pauli matrices for spins located at site $i$, $P$ refers to a square plaquette, and $ijkl\in P$ to the four sites at the corners of the plaquette.
The first term is the four spin plaquette interaction, while the second term is the external transverse field. 
While the conventional Ising model contains only a global $Z_2$ symmetry, the PIM contains subextensively many $d=1$ subsystem $Z_2$ symmetries.
These symmetries corresponding to flipping all spins $\sigma_i^z\rightarrow-\sigma_i^z$ along any row or column, which leave the Hamiltonian invariant.
We therefore have $L_x+L_y-1$ unique $Z_2^{sub}$ symmetry operators, where the superscript serves as a reminder that we are dealing with subsystem symmetry, and the $-1$ comes from the fact that flipping all columns is the same as flipping all rows.

For small $\Gamma$ and zero temperature, this model enters a spontaneous symmetry broken ordered phase where all spins align such that every plaquette term in the Hamiltonian is satisfied.  
The ground state is $2^{L_x+L_y-1}$-fold degenerate, and consists of spin states related to the trivial $z$-polarized state by applications of the subsystem symmetry.

In the opposite limit, $\Gamma \gg 1$, the ground state is the unique paramagnetic phase with all spins polarized $\sigma_i^x=1$.
In the $z$-basis, such a state is an equal superposition of all possible configurations of $\sigma^z_i$.
The paramagnetic ground state of the PIM contains no entanglement, and will sometimes be referred to as the topologically trivial paramagnet.
We now describe two distinct paramagnetic phases protected by the $Z_2^{sub}$ symmetry - these are our first examples of SSPTs.

\subsection{Weak SSPT}

We first illustrate the construction of a ``weak'' SSPT phase.  
Such phases may be adiabatically continued to a state consisting of decoupled $1D$ SPT chains without closing the gap or breaking any of the subsystem symmetries.

\subsubsection{$1D$ $Z_2\times Z_2$ SPT}\label{sss:1dspt}
First, we review the $1D$ cluster Hamiltonian, whose ground state describes an SPT phase protected by a global $Z_2\times Z_2$ symmetry~\cite{chen2014symmetry}.
We take a chain and label the two sublattices $A$ and $B$.
For each site $i$ on the $A$ ($B$) sublattice, we have a spin-$1/2$ degree of freedom on which the Pauli matrices $\sigma_i^{x,y,z}$ ($\tau_i^{x,y,z}$) act.
The Hamiltonian is given by
\begin{equation}
    H_\text{1d} = -\sum_{i\in A} \tau^z_{i-1} \sigma^x_{i} \tau^z_{i+1} -\sum_{i\in B} \sigma^z_{i-1} \tau^x_{i} \sigma^z_{i+1}
\label{chain}
\end{equation}
This system possesses a global $Z_2 \times Z_2$ symmetry which consists of flipping all $\sigma^z$ or all $\tau^z$ spins and is generated by the operators $\prod_{i\in A}\sigma^x_i$ and $\prod_{i\in B}\tau^x_i$.  
The ground state is $Z_2\times Z_2$ symmetric and in the $\{\sigma^z_i, \tau^x_i\}$ basis, it is an equal superposition of all possible $\{\sigma^z_i\}$, but with domain walls $\sigma^z_{i-1}\sigma^z_{i+1}=-1$ decorated by a $\tau^x_{i}=-1$ ($\tau^x=+1$ otherwise).
As there is one term in the Hamiltonian that must be satisfied per site, this ground state is unique for periodic boundary conditions. 

One can see that this phase belongs to a non-trivial topological phase by observing that introducing a boundary produces a $2$-fold degeneracy that cannot be broken (while preserving the symmetry).  
Furthermore, the action of the symmetry localized at one edge realizes a projective representation of the symmetry group $Z_2\times Z_2$.

Let us consider an open system of length $L$ and suppose that both edges are terminated by a $\sigma$ spin (for demonstration purposes).  
We also exclude any term in the Hamiltonian that is not fully contained in the system, to ensure that no symmetry is broken.
Notice however, that there are now only $L-2$ terms in the Hamiltonian, while there are $L$ spins, and so we now have a $2^2$-fold degeneracy ($2$ from each edge).  
We may define two sets of Pauli matrices located at the left and right edges,
\begin{eqnarray} \label{1dPis}
    \pi^z_{l} = \sigma_{1}^z&,&\; \pi^{x,y}_{l} = \sigma_{1}^{x,y}\tau_{2}^{z}\\
    \pi^z_{r} = \sigma_{L}^z&,&\; \pi^{x,y}_{r} = \tau_{L-1}^{z}\sigma_{L}^{x,y}
\end{eqnarray}
which obey the Pauli algebra and commute with every term in the Hamiltonian. 

It is straightforward to show that
\begin{eqnarray} \label{1dSyms-1}
    \prod_{i\in A}\sigma^x_i &=& \pi^x_l \pi^x_r \\
    \label{1dSyms-2}
    \prod_{i\in B}\tau^x_i &=& \pi^z_l \pi^z_r 
\end{eqnarray}
on the ground state manifold, using the fact that the ground states are eigenstates of every term in the Hamiltonian.
Thus, {\it the action of the symmetries can be factored into operations acting on the left and right edges separately.}

When the global symmetry factors in this way, it is possible for the symmetry at one boundary (say the left one) to act {\it projectively}, i.e. with phases that are not present in the action of the global symmetry itself.\cite{Pollmann2012-lv,Chen2011-et}   Such phases could arise by simply re-defining the symmetry action at the left boundary by a phase (say $\pi^x_r \rightarrow e^{i \alpha}\pi^x_r$), and the symmetry action at the right boundary by the conjugate phase $\pi^x_l \rightarrow e^{ - i \alpha}\pi^x_l$).  Such arbitrary phase factors clearly do not tell us anything about the underlying physics, and are not associated with true projective representations.  However, Eqs. (\ref{1dSyms-1}- \ref{1dSyms-2}) exhibit a different type of phase, since at a given edge, the operators associated with global $\sigma$ spin flips and global $\tau$ spin flips anti-commute.  The resulting phase cannot be eliminated by the phase choice described above (which will only move it from one symmetry process to another).  Thus this anti-commutation indicates 
 that the symmetry group $Z_2\times Z_2$ is realized \emph{projectively} at each edge.

To see how this projective nature protects the boundary degeneracy, suppose we add arbitrary perturbations that do not break any symmetry.  
We may always project on to the low energy subspace of $H_{1d}$ to observe how the perturbation acts on the low energy manifold.
Any perturbation localized on the left edge cannot break the degeneracy, as it must commute with both $\pi^x_l$ and $\pi^z_l$ (and similarly for the right edge).  
In order to break the degeneracy in the thermodynamic limit, one must either break the symmetry or introduce a nonlocal perturbation (or undergo a bulk phase transition).

\subsubsection{$Z_2^{sub}$ Weak SSPT}
To construct a ``weak'' SSPT phase,
let us align stacks of $1D$ $Z_2\times Z_2$ SPTs previously discussed along both the $x$ and $y$ directions, such that each site of the resulting square lattice contains two $\sigma$ or two $\tau$ spins from two intersection $1D$ chains.
We then consider the whole $2D$ system, and call the $Z_2\times Z_2$ symmetry of each individual chain our subsystem symmetries, such that our total symmetry group is now $(Z_2^{sub}\times Z_2^{sub})^{N_\text{chains}}$, where $N_\text{chains}$ is the total number of $1D$ chains in our system.
The chains may then be coupled weakly in a way that respects all the subsystem symmetries.

Now suppose we have an open system with dimensions $L$ and boundaries along the $x$ or $y$ direction.  
Each SPT chain that is cut produces a $2$-fold degeneracy at its end.  
Thus, our system as a whole has a subextensive ground state degeneracy, growing as $2^{\mathcal{O}(L)}$, that cannot be broken with local symmetry-respecting perturbations.
Next, to consider the projective representation of the symmetry at an edge, consider a boundary along the $x$ direction on which $2\ell$ subsystems terminate (2 for each of $\ell$ columns).   
As discussed in the previous section, we have a projective representation of $Z_2^{sub}\times Z_2^{sub}$ for each column.

We should note that the microscopic action of the symmetry in this weak SSPT is fundamentally different from the 2D strong SSPT to be described next.  
Although the subsystem symmetry along the rows and columns overlap spatially, they act on distinct physical spins.  
Thus, although each \emph{site} is acted on by two different subsystem symmetries, each \emph{spin} is only flipped by one.  
In the strong SSPT to be introduced, as well as in the trivial plaquette paramagnet, each spin shall be flipped by two distinct symmetries.

\subsection{$Z^{sub}_2$ Strong SSPT}

\begin{figure}[h]
  \centering
      \includegraphics[width=0.25\textwidth]{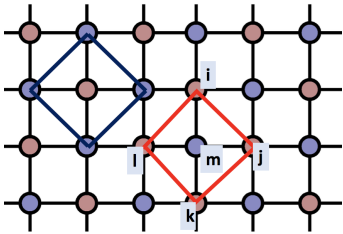}
  \caption{The terms in the TPIM Hamiltonian. The Pauli spins $\tau,\sigma$ live on the red/blue sites. The interaction $ \sigma^z_i \sigma^z_j \sigma^z_k \sigma^z_l  \tau^x$ involves the four $\sigma_z$ spins on the blue plaquette and the $\tau_x$ in the middle. The interaction $\tau^z_i \tau^z_j \tau^z_k \tau^z_l  \sigma^x$ involves the four $\tau_z$ spins on the red plaquette and the $\sigma_x$ in the middle. }
  \label{one}
\end{figure}

\begin{figure}[h]
  \centering
      \includegraphics[width=0.24\textwidth]{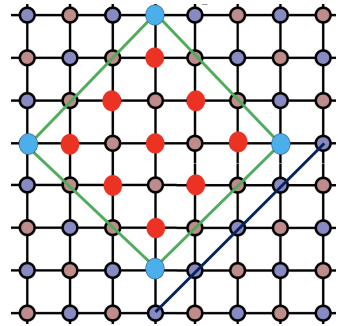}
  \caption{The dark blue line gives an example of a subsystem: a single row on sublattice $A$ (where we call the spin operator $\sigma$). The  green square indicates the boundary of the membrane order parameter, which involves product of $\sigma_z$(blue) at the corner of the membrane and product of $\tau_x$(red) inside the membrane.}
  \label{two}
\end{figure}

Next, we study a 2D cluster Hamiltonian introduced by Ref.~\cite{Raussendorf2001-xm}, and show that it realizes a strong SSPT, which we refer to as topological plaquette Ising model (TPIM).
The Hilbert space consists of Ising spins on sites of the square lattice.  For clarity, we will separate these into two spin flavours, $\sigma$ and $\tau$, located at the sites of the $A$ and $B$ sublattices, respectively.
The Hamiltonian is given by

\begin{align} 
H_\text{TPIM}=-\sum_{ijklm \in P_A} \sigma^z_i \sigma^z_j \sigma^z_k \sigma^z_l  \tau^x_m-\sum_{ijklm \in P_B} \tau^z_i \tau^z_j \tau^z_k \tau^z_l  \sigma^x_m
\label{topo1}
\end{align}
where the sum is over all $P_A$ ($P_B$), which refer to five-site clusters consisting of a site on the $A$ ($B$) sublattice and its four nearest neighbors, with each site labeled by $ijklm$ as illustrated in Fig~\ref{one}.
The first term is a sum over products of a $\tau^x$ and its four surrounding $\sigma^z$, and vice versa for the second.
As all local cluster-operators commute with each other, the Hamiltonian contains extensively many conserved quantities and is exactly solvable.
Indeed the ground state of this Hamiltonian is the well studied 2D Cluster state on the square lattice~\cite{Raussendorf2001-xm}. 
In addition, the model has $Z^{sub}_2 $ symmetry, as the Hamiltonian commutes with the operators $\prod_{\text{diag}} \sigma^x$ and $\prod_{\text{diag}} \tau^x$ which flips $\sigma_z \rightarrow -\sigma_z$ or $\tau_z \rightarrow -\tau_z$ along a particular diagonal (see Fig~\ref{two}).

\begin{figure}[h]
  \centering
      \includegraphics[width=0.4\textwidth]{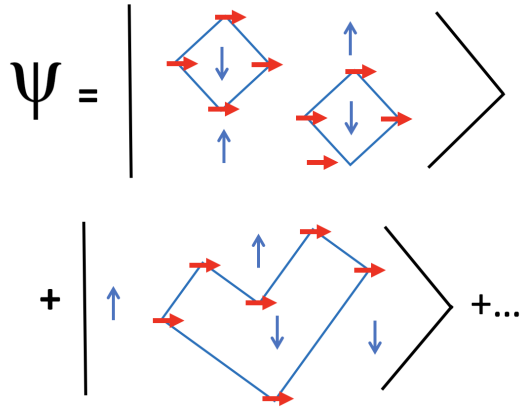}
  \caption{
  Ground state of the TPIM. The blue lines denote domain walls for $\sigma$ spins, where $\sigma_z= +1 (-1)$ outside/inside a domain. 
  The corners of these domains are decorated by a $\tau^x=-1$ spin, indicated by the red arrows. 
  The ground state is an equal superposition of all such configurations.
  }
  \label{ten}
\end{figure}

To understand the ground state of this Hamiltonian, we work in the $\sigma^z$ and $\tau^x$ basis.  
The first term in the Hamiltonian means that $\tau^x_m = \sigma^z_i\sigma^z_j\sigma^z_k\sigma^z_l$ for $ijklm\in P_a$.
That is,  plaquettes where the product $\sigma^z_i\sigma^z_j\sigma^z_k\sigma^z_l = -1$ are decorated with $\tau^x=-1$, otherwise $\tau^x=+1$ (such plaquettes appear at the \emph{corner} of a domain wall, as illustrated in Fig~\ref{ten}).
The second term in the Hamiltonian flips a single $\sigma^z$, and the surrounding $\tau^x$ appropriately, transitioning between two valid configurations.
Thus, the ground state of $H_\text{TPIM}$ can be described as a superposition of all possible $\{\sigma^z\}$ configurations, with the corners of each domain wall decorated with $\tau_x=-1$, as shown in Fig.~[\ref{ten}].  This is similar to the decorated defect construction for $2D$ global SPT phases \cite{chen2014symmetry,you2016stripe,you2016decorated}. 

Finally, we note that the TPIM Hamiltonian can be perturbed with a subsystem symmetry-preserving term, the simplest of which is an on-site transverse field:
\begin{equation}
    H = H_\text{TPIM} - \Gamma \sum_{i\in a}\sigma^x_i - \Gamma \sum_{i\in b}\tau^x_i .
\end{equation}
As we show in Appendix \ref{app:duality}, this Hamiltonian admits a duality transformation to two copies of the PIM, whith the SSPT phase being mapped to the phase with spontaneously broken subsystem symmetry. The latter is known to have a $4^{L_x + L_y -1}$-fold degeneracy due to the $4^{L_x + L_y -1}$ spontaneously broken Ising symmetries. We will see presently that our SSPT phase has the same degeneracy, resulting from gapless boundary modes.  
As discussed in the Appendix the model is also self-dual, with $\Gamma \leftrightarrow \Gamma ^{-1}$.
%and can also be mapped to  (although with different degeneracies) to the 2D quantum compass model~\cite{Kugel1982-ui} and the Kitaev toric code in a transverse field~\cite{Vidal2009-lc}.
From these mappings, we learn that when the perturbation reaches $\Gamma =1$, the model exhibits a first order transition~\cite{Orus2009-zh,Kalis2012-io} to (two copies of) the trivial $Z^{sub}_2$ paramagnet.

While there is no local order parameter for distinguishing the TPIM and PIM ground states, 
there exists a string order parameter~\cite{Doherty2009-fd}, which can be straightforwardly generalized 
to a fully two-dimensional membrane order parameter $O$,
\begin{align} 
O=\langle \prod_{i \in \mathcal{C}}  \sigma^z_i  \prod_{i \in \mathcal{M}} \tau^x_i   \rangle
\label{mem}
\end{align}
Here $\mathcal{C}$ refers to the $A$ sites on the corners of the membrane and $\mathcal{M}$ contains all $B$ sites inside the membrane, as depicted in Fig~\ref{two}.
Taking the membrane size to infinity, this order parameter approaches a constant in the SSPT phase, and zero in the trivial   subsystem symmetric phase. 
This nonlocal membrane operator captures the decoration of the domain wall corners and can serve as a numerical signature to detect the topological plaquette paramagnet.

\subsubsection{Edge states}\label{sss:edgestates}
A distinctive feature of SPT states in 1 and 2 dimensions is the existence of gapless symmetry-protected boundary modes, which cannot be gapped unless the global symmetry is broken~\cite{Kane2005-ml,Chen2012-oa,Vishwanath2013-pb}. 
Here we show that the SSPT paramagnet similarly has non-dispersing gapless boundary modes protected by the subsystem symmetry, which leads to a subextensive ground state degeneracy in the presence of an edge.

Our argument will proceed as follows.  
We first consider the system with an edge, and simply omit terms in $H_\text{TPIM}$ that are not fully contained in the bulk of the system.
Looking at what lives on the edge, we find that this leaves a free spin-$1/2$ degree of freedom per unit length along the boundary.
We then ask whether other local terms commuting with the symmetries (whether or not they commute with $H_\text{TPIM}$) can be added to lift this degeneracy along the edge, to which the answer is no.
We therefore conclude that this system has a symmetry protected $2$-fold degeneracy per unit length along the edge.

\begin{figure}[h]
  \centering
      \includegraphics[width=0.25\textwidth]{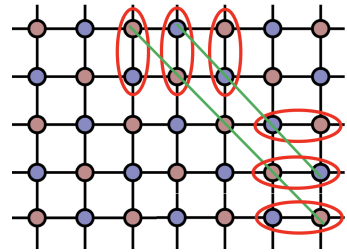}
  \caption{
  Red ovals show the physical spins that take part in the edge operators $\pi^\alpha_i$, and form a spin-$1/2$ degree of freedom at the edge.  
  The action of the subsystem symmetries (green lines) on the ground state manifold may be expressed in terms of such $\pi^\alpha_i$ operators.  
  Near a corner of the type shown here, the symmetry becomes a local symmetry, and the corresponding boundary modes can be gapped out.}
  \label{three}
\end{figure}

First, let us consider a horizontal/vertical edge as shown in Fig.~[\ref{three}]. 
Along this edge, we may pick two-spin clusters (red ovals in Fig~\ref{three}), which each contain a $\tau$ and a $\sigma$ spin.
These spin clusters create a free spin 1/2 degree of freedom on each site along the edge. 
To see this, %we can define the new edge operators\cite{bahri2015localization,Chen2011-et}:
observe that for each edge site with a $\tau$ spin at the surface, we have the three edge operators
\begin{align} 
\pi^x=\sigma^z \tau^x,\pi^y=\sigma^z \tau^y, \pi^z= \tau^z
\label{edge1}
\end{align}
and likewise, for odd edge sites with $\sigma$ spin at the surface, we have
\begin{align} 
\pi^x=\tau^z \sigma^x,\pi^y=\tau^z \sigma^y, \pi^z= \sigma^z
\label{edge2}
\end{align}
As for the 1D $Z_2 \times Z_2$ SPT \cite{bahri2015localization,Chen2011-et}, these operators satisfy the Pauli algebra on the surface, and commute with the bulk Hamiltonian $H_\text{TPIM}$.
By counting degrees of freedom, we can see that there exists a $2^L$-fold degenerate ground state manifold arising due to the presence of the edge of length $L$, which these $L$ Pauli operators act on.

This edge degeneracy in fact cannot be broken while preserving all subsystem symmetries, and leads to a completely flat-band dispersion along the edge.  
To see this, we may use the same argument as we used before for the $1D$ SPT,
and consider the action of the subsystem symmetries on the ground state manifold in terms of these $\pi^\alpha_i$ operators.
Considering only the action on a single edge, for each site $i$ along the edge, there exist two symmetries which act as $S_i^{(1)} = \pi_i^{z} \pi_{i+1}^x$ and $S_i^{(2)} = \pi_i^{x} \pi_{i+1}^z$ along that edge.

Notice that in our description of the low-energy Hilbert space at a single edge,  there are neighboring symmetry operators ($S_i^{(1)}$ and $S_{i+1}^{(1)}$ for example) do not commute.  
This is an artefact of restricting our attention to a single edge at a time: the full symmetry acts simultaneously on pairs of edges of the system, such that the symmetry operators applied to the system as a whole do commute.  
However, much as for the AKLT chain\cite{Pollmann2012-lv} and our $1D$ SPT earlier, this apparent non-commutativity reflects the fact that the symmetry group is realized projectively at the boundary.
While the form of these edge operators will depend on our definition of $\pi^\alpha_i$ and the microscopic details of the edge cut, their non-commutativity is independent of such details (to see this, notice that one is free to make any type of cut at the \emph{other} edge, and that the symmetry as a whole is realized linearly). 

We may then ask whether terms may be added to the Hamiltonian that can break the degeneracy of the ground state manifold (away from the corners).
Any term which we add to the Hamiltonian respecting all symmetries, projected to the degenerate subspace perturbatively via the effective Hamiltonian, must still commute with all symmetries in the effective Hamiltonian.
It is easy to see that no local (non-identity) term can be written down along this edge which commutes with all $S_i^{(1,2)}$, and therefore the effective Hamiltonian along this edge must be trivially proportional to identity.  
Indeed any state that respects all of these symmetries must have a 2-fold degeneracy per unit length along the edge.

Near $90^\circ$ corners of this type, however, a gap may be opened. 
This can be seen by noting that some subsystem symmetries (which go diagonally) essentially become local symmetries near the corners as Fig.~[\ref{three}]. Thus, the symmetries themselves and products thereof (which commute with all other symmetries and are local near the corners) may be included as terms in the effective Hamiltonian, thus lifting the exact degeneracy.  

\begin{figure}[h]
  \centering
      \includegraphics[width=0.3\textwidth]{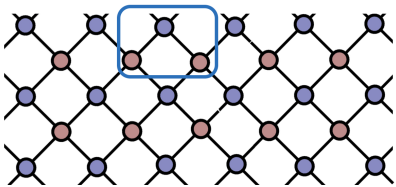}
  \caption{The spins in the blue rectangle are involved in the edge operators $\pi^\alpha_i$, in the case of a $45^\circ$ edge. } 
  \label{four1}
\end{figure}

A similar argument applies for an edge cut along the $45^\circ$ direction.
The edge $\pi$ degrees of freedom are now composed of three spin clusters, depicted in Fig.~[\ref{four1}], given by
\begin{align} 
\pi^x=\tau^z \sigma^x \tau^z,\pi^y=\tau^z \sigma^y \tau^z, \pi^z= \sigma^z
\label{edge3}
\end{align}
for the cluster with a $\sigma$ at the edge (and similarly for the clusters with $\tau$ on the edge, with $\sigma\leftrightarrow\tau$).
Similar to the earlier cut, there are two symmetries per site along the edge which act in the effective edge Hamiltonian as
$S_{i}^{(1)} = \pi_i^x$ and $S_{i}^{(2)} = \pi_i^{z} \pi_{i+1}^{z}$, along with the symmetry that acts globally along the edge as $\prod_{i} \pi_i^x$.
As before, there are no local terms that can be added to the Hamiltonian respecting all symmetries, and thus there is a degeneracy along this edge protected by the subsystem symmetries.

\subsection{Distinctions between weak and strong SSPTs}\label{distinct}

At this stage, we would like to comment on the differences between the weak SSPT phase obtained from stacking $1D$ SPTs, and the strong $2D$ SSPT. 
We note that in the two explicit models discussed above the subsystem symmetries are different: in the weak SSPT each spin is flipped {\it either} by a horizontal or by a vertical subsystem symmetry.  In the strong SSPT, in contrast, each spin is flipped by {\it both} a horizontal and a vertical subsystem symmetry.  
Since the symmetries in these two models differ, 
prima facie there can be no path between the two ground states that preserves all symmetries.

A more subtle question is whether the TPIM is intrinsically distinct from the weak SSPT phase, or whether the difference noted above is an artefact of our particular construction.
In 1D, it is known \cite{Turner2011-zi,Fidkowski2011-be,Chen2011-et} that two distinct phases with the same unbroken symmetry realize different projective representations of this symmetry at their boundaries.
Briefly reviewing the 1D SPT with global $Z_2\times Z_2$ symmetry from Sec~\ref{sss:1dspt}, we found that the symmetry action on the ground state manifold, may be decomposed into operators acting on the left and right edges separately, which inevitably anticommuted among themselves.
These two sets of operators therefore generate a projective representation of $Z_2\times Z_2$, which can be characterized by these anticommuting operators.
Returning to our prior discussion, we may ask what projective representation of $Z_2^{sub}$ is realized along the edge of our 2D system of weakly coupled 1d SPT chains.  
Here we find that the edge action of the generators of our total symmetry group $(Z_2)^{N_{sub}}$, where $N_{sub}$ is the total number of subsystems terminating along the edge, can be decomposed into $N_{sub}/2$ pairs which each locally form the projective representation of $Z_2\times Z_2$ described above.

For the strong SSPT however, we find that  the edge action of the symmetries is quite different.
In Section~\ref{sss:edgestates}, we found that the symmetries acting on the edge, in terms of edge degrees of freedom $\pi^\alpha_i$, behaved as $\pi^x_i \pi^y_{i+1}$ and $\pi^y_i \pi^x_{i+1}$ for a vertical/horizontal cut, or as $\pi^x_i$ and $\pi^z_i \pi^z_{i+1}$ for a diagonal cut. 
In either case, we notice that each symmetry operator anticommutes with $\emph{two}$ neighboring operators. This is in contrast with the boundary of the weak SSPT, where each operator anticommuted with only one other.

Indeed, the intertwining pattern of anti-commutators ensures that the projective representation describing the boundary of a strong SSPT with $N$ sites corresponds to a projective representation of $Z_2^{N}$ that cannot be expressed in terms of copies of a projective representation of $Z_2^m$ for any $m < N$. 
In other words, the boundary of the strong SSPT phase cannot be obtained from decoupled 1D SPTs.  In Appendix~\ref{app:groupcoho}, we present a more general calculation indicating for which subsystem symmetry groups such projective representations exist, guaranteeing that the weak and strong SSPT phases have different symmetry realizations at their boundaries.  Notably, we find that such representations do {\it not} exist for familiar continuous symmetry groups such as U(1), SU(2), or SO(3).

The projective representation realized along the edge is therefore obviously distinct from that of the decoupled chains.
Thus, our strong $Z_2^{sub}$ SSPT must exist as a distinct phase from any weak SSPT with the same symmetries.

\subsubsection{Response to flux insertion}
\begin{figure}[h]
  \centering
      \includegraphics[width=0.3\textwidth]{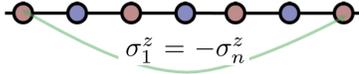}
  \caption{The SPT chain with anti-periodic boundary conditions in $\sigma^z$ (and periodic boundary conditions in $\tau^z$). $tau^z$ is identified on the two sites connected by the green line, while $sigma^z$ changes sign as the green line is crossed. } 
  \label{flux1}
\end{figure}

We now turn to a different approach to distinguish the weak and strong SSPT phases -- via their response to flux insertion.
For the 1$D$ $Z_2\times Z_2$ SPT chain in Eq.~[\ref{chain}], one can gauge one of the $Z_2$ symmetries by coupling the Ising spin with a $Z_2$ gauge connection $\Pi^z=e^{i A_x}$ living on the link between two nearest $\sigma$ spins. 
\begin{align} 
\sigma^z_{i} \sigma^z_{i+1} \tau^x_{i+1/2} \rightarrow \sigma^z_{i} \Pi^z_{i, i+1} \sigma^z_{i+1} \tau^x_{i+1/2}
\end{align}
We now place the SPT chain on a ring and make a large gauge transformation by inserting a $\pi$ gauge flux through the ring \cite{santos2014symmetry} by requiring: 
\begin{align} 
e^{i \int dx A_x}=\prod_i \Pi^z_{i, i+1}=-1 \ .
\end{align}
The flux insertion imposes an anti-periodic boundary condition $\sigma^z_1=-\sigma^z_n$, as shown in Fig.~[\ref{flux1}]. 
Periodic boundary conditions enforce an even number of domain walls for $\sigma^z$ along the chain, so that the total $Z_2$ charge due to the $\tau^x$ spins decorating the domain walls along the chain is also even. Once we impose anti-periodic boundary conditions in $\sigma^z$, there are an odd number of domain walls along the chain.  In this case the decorating charge, which we can measure via the charge parity operator,
\begin{align} 
&L= e^{i \pi [\sum_i (1-\tau^x_i)/2]}
\end{align}
is also odd.

For the weak SSPT built from aligned 1$D$ SPT chains, we can gauge the `subsystem $Z_2$ symmetry' by imposing anti-periodic boundary conditions for a specific chain.  (Since each chain has its own $Z_2\times Z_2$ symmetry, here it makes sense to consider changing the boundary conditions of the chains individually).  This would change the $\tau^x$- charge parity of only the affected chain.

\begin{figure}[h]
  \centering
      \includegraphics[width=0.5\textwidth]{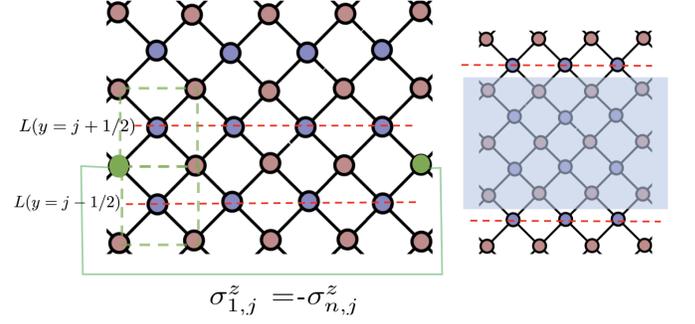}
  \caption{L: Imposing anti-periodic boundary conditions for $\sigma^z_{1,j}=-\sigma^z_{n,j}$（green sites) by inserting higher rank gauge flux. This switches the charge parity $L$ (red dashed lines) on the rows at $y=j\pm 1/2$, and effectively switches the sign of two plaquette terms in the Hamiltonian (indicated by the green dashed squares ).  R: Imposing anti-periodic boundary conditions inside the blue shaded membrane region. This switches the charge parity $L$ (red dashed lines) on the rows at the boundary of the membrane.} 
  \label{flux2}
\end{figure}

Now we turn to the case of the strong $Z^{sub}_2$ SSPT state. To gauge the part of $Z^{sub}_2$ associated with $\sigma$, we are led to introduce a rank-2 gauge connection $\Pi^z=e^{iA_{xy}}$ in the center of each plaquette \cite{pretko2017fracton,Pretko2017-nt,Pretko2017-nt,Vijay2015-jj,Vijay2016-dr}, and couple the gauge connection with the four spins on the plaquette via
\begin{align} 
&\sigma^z_{i,j} \sigma^z_{i,j+1}\sigma^z_{i+1,j}\sigma^z_{i+1,j+1} \tau^x_{i+\frac{1}{2},j+\frac{1}{2}}\nonumber\\
&\rightarrow \sigma^z_{i,j} \sigma^z_{i,j+1}\sigma^z_{i+1,j}\sigma^z_{i+1,j+1} \tau^x_{i+\frac{1}{2},j+\frac{1}{2}} \Pi^z_{i+\frac{1}{2},j+\frac{1}{2}}  \ 
\end{align}
%where the gauge connection associated with $\sigma$ spins (which live on the $A$ sublattice) lives on the sites of the $B$ sublattice.
Consider placing the SSPT state on a cylinder with periodic boundary conditions in $x$ and open boundary conditions along $y$.  The analogue of flux insertion for our rank-2 gauge connection is to insert $\pi$ flux between the $j-1/2$-th and $j+1/2$-th rows,
\begin{align} 
B^y(j\pm 1/2)=e^{i \int d_x A_{xy}}=\prod_{i}\pi^z_{i+\frac{1}{2},j\pm \frac{1}{2}} =-1
\end{align}
This imposes anti-periodic boundary conditions for the $j$-th row, $\sigma^z_{1,j}=-\sigma^z_{n,j}$, as in Fig.~[\ref{flux2}]. Meanwhile, other rows still have periodic boundary conditions. 
This effectively changes the sign of the two plaquette terms containing site sites $1,j$ and $n,j$ (indicated by the green dashed squares in Fig.~[\ref{flux2}]):
\begin{align} 
&\sigma^z_{1,j} \sigma^z_{1,j+1}\sigma^z_{n,j}\sigma^z_{n,j+1} \tau^x_{\frac{1}{2},j+\frac{1}{2}}\nonumber\\
&\rightarrow -\sigma^z_{1,j} \sigma^z_{1,j+1}\sigma^z_{n,j}\sigma^z_{n,j+1} \tau^x_{\frac{1}{2},j+\frac{1}{2}}\nonumber\\
&\sigma^z_{1,j} \sigma^z_{1,j-1}\sigma^z_{n,j}\sigma^z_{n,j-1} \tau^x_{\frac{1}{2},j-\frac{1}{2}}\nonumber\\
&\rightarrow -\sigma^z_{1,j} \sigma^z_{1,j-1}\sigma^z_{n,j}\sigma^z_{n,j-1} \tau^x_{\frac{1}{2},j-\frac{1}{2}}  \ \ .
\end{align}

With periodic boundary conditions, the system has an even number of domain wall corners along each row/column, and the total $\tau$ charge along each row/column is even as well. Imposing the anti-periodic boundary condition at a specific row $\sigma^z_{1,j}=-\sigma^z_{n,j}$ changes the number of domain wall corners in rows $j \pm 1/2$ (see Fig.~[\ref{flux2}]) from even to odd, which also switches the parity of the $\tau$ charge, defined by:
\begin{align} 
&L(y=j \pm 1/2)= e^{i \pi [\sum_i (1-\tau^x_{i,j \pm 1/2})/2]}
\end{align}

Thus for the weak SSPT, we find that twisting the boundary condition in a single row leads to a response in that row, while for the strong SSPT, we see a response in a pair of adjacent rows.  
This difference in response of the weak and strong SSPT phases can be seen more clearly if we apply anti-periodic boundary conditions to all rows in a finite-width strip, as shown in Fig.~[\ref{flux2}]. For the strong SSPT, this alters the gauge field configuration only on the border of the membrane, and switches the charge parity only in the corresponding two rows (red dashed lines in Fig.~[\ref{flux2}]).  For the weak SSPT state, however, this operation changes the $tau$-charge parity on every row inside the membrane. This charge parity response under twisted boundary conditions could be used as a computational identification of the SSPT phase.

\subsection{$Z^{sub}_n \times Z^{sub}_m$ SSPT phases}

Besides $Z_2^{sub}$, there are also other subsystem symmetry groups for which a strong SSPT phase exists and we get projective representations for a boundary which cannot be generated from copies of projective representations for smaller systems.
In Appendix~\ref{app:groupcoho} we show that this is possible in general for discrete abelian groups others, such as $Z_n$ or $Z_n \times Z_m$, because a certain torsion term in their group cohomology is non-vanishing.  (The relevant torsion vanishes for familiar continuous groups, such as $SO(2)$, $SO(3)$, or $U(1)$), and we do not know of a model realizing $d=1$ SSPT phases for these).   
Here, we provide an explicit construction for one such strong SSPT phase, which is protected by $Z_n^{sub}\times Z_m^{sub}$ symmetry.

We replace $\tau$ and $\sigma$ spins by $n$ and $m$ dimensional degrees of freedom, on which we introduce
local $Z_n$ operators $\mathrm{Z}$ and $\mathrm{X}$ satisfying
\begin{eqnarray}
    \mathrm{X}^n = \mathrm{Z}^n = 1\\
    \mathrm{XZ} = \omega \mathrm{ZX}
\end{eqnarray}
with $\omega=e^{2\pi i/n}$, and similarly $Z_m$ operators $\mathrm{\tilde{Z}}$ and $\mathrm{\tilde{X}}$ with $\tilde{\omega} = e^{2\pi i/m}$.

Then, assuming $n$ and $m$ have a nontrivial greatest common divisor $q=\gcd(a,b)\neq 1$, one can write the Hamiltonian
\begin{eqnarray}
    H &=& -\sum_{ijklm\in P_a} (\mathrm{\tilde{Z}}_i\mathrm{\tilde{Z}}^\dagger_j\mathrm{\tilde{Z}}_k\mathrm{\tilde{Z}}^\dagger_l)^{\frac{mz}{q}} \mathrm{X}_m + h.c. \nonumber \\
    &&-\sum_{ijklm\in P_b} (\mathrm{Z}_i\mathrm{Z}^\dagger_j\mathrm{Z}_k\mathrm{Z}^\dagger_l)^{\frac{nz}{q}} \mathrm{\tilde{X}}_m
     + h.c.
\end{eqnarray} 
for any integer $z$
which consists of mutually commuting terms and is therefore exactly solvable.
One may verify that each choice of $z=1\dots q$ corresponds to a different projective representation of the subsystem symmetries along the edges.

\section{3D topological cubic paramagnetic phase} \label{3dversion}
In this Section, we show how a $d=1$ subsystem symmetry can lead to new SSPT phases in 3 dimensions. We illustrate this by constructing an exactly solvable Hamiltonian with $Z_2^{sub}$ subsystem symmetry.  As for our 2-dimensional model above, this model has symmetry-protected non-dispersing gapless boundary modes, an entangled ground state, and can be detected via a non-local order parameter.  

As before, we start with the trivial cubic paramagnet, given by the Hamiltonian
\begin{align} 
&H_{\text{CIM}}=-\sum_{C} \prod_{i\in C}\sigma^z_i  -h\sum_{i} \sigma_i^x
\label{cising}
\end{align}
which we refer to as the Cubic Ising model (CIM).
The $\sigma$ spins lie on the sites of a simple cubic lattice.
The sum over $C$ sums over cubes, and $i\in C$ refers to the 8 spins on the same cube.
The first term involves 8-site interaction on a cube and the second term is the transverse field. 
This Hamiltonian contains $d=1$ subsystem $Z_2^{sub}$ symmetries, which involve flipping $\sigma_z \rightarrow -\sigma_z$ along a line in either the $x$,$y$, or $z$ direction.
There are 
\begin{equation}
    \mathcal{D}=L_xL_y+L_yL_z+L_zL_x-L_x-L_y-L_z-2
\end{equation} independent operators on an $L_x L_y L_z$ torus.
For $h\ll 1$, the ground state spontaneously breaks these symmetries and is $2^\mathcal{D}$-fold degenerate, and for $h>>1$, the system is in its trivial paramagnetic phase with $\sigma_i^x=1$.

\begin{figure}[h]
  \centering
      \includegraphics[width=0.3\textwidth]{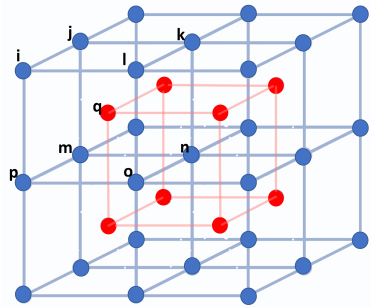}
  \caption{
  The BCC lattice on which the TCIM is defined.  Spin-$1/2$ degrees of freedom $\sigma$ ($\tau$) live on the blue (red) sublattice, each of which form their own simple cubic lattice.
  The spin interactions involve the eight spins on a cube of one sublattice and one from the other.
  } 
  \label{six}
\end{figure}

We can now create the topological cubic paramagnetic state by condensing appropriately decorated domain surfaces, similar to our construction in 2$d$.
The resulting model can be regarded as the cluster Hamiltonian on the body centered cubic (BCC) lattice.
The BCC lattice can be regarded as two displaced simple cubic lattices, labeled by the blue/red sites in Fig.~[\ref{six}], which we call the $A$ and $B$ sublattices, respectively.
Each lattice site contains a spin-1/2 degree of freedom, and for convenience we label the spins on the blue sites $\sigma$, and those on the red sites $\tau$.
The Hamiltonian is given by
\begin{align} 
H_\text{TCIM} =&-\sum_{ijklmnopq \in C_A}  \sigma^z_i \sigma^z_j \sigma^z_k \sigma^z_l\sigma^z_m \sigma^z_n \sigma^z_o \sigma^z_p \tau^x_q \nonumber\\
& -\sum_{ijklmnopq \in C_B} \tau^z_i \tau^z_j \tau^z_k \tau^z_l \tau^z_m \tau^z_n \tau^z_o \tau^z_p \sigma^x_q
\label{cubic}
\end{align}
Here $C_A$ ($C_B$) refers to a site on the $A$ ($B$) sublattice and its eight nearest neighbors, labeled by $ijklmnopq$ as depicted in Fig.~[\ref{six}].
This Hamiltonian is again composed of commuting terms and is therefore exactly solvable.
The subsystem symmetry in this case corresponds to flipping all $\sigma^z$ or $\tau^z$ spins along a line in the $x$,$y$, or $z$ direction, which we implement as $\prod_{i\in \text{line}} \sigma^x_i$ or $\prod_{i\in \text{line}} \tau^x_i$.

\begin{figure}[h]
  \centering
      \includegraphics[width=0.3\textwidth]{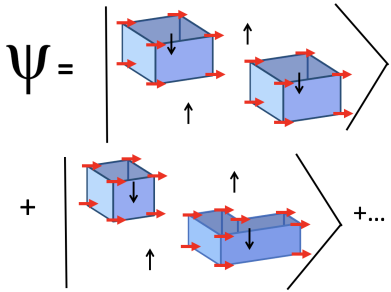}
  \caption{Ground state of the TCIM as a superposition of all domain wall configurations for $\sigma$ spins. The blue blocks represent domain walls; $\sigma_z$ changes from $+1$ to $-1$ across the domain wall. Each corner of the domain wall is decorated with $\tau_x=-1$.} 
  \label{11}
\end{figure}

The ground state wave function is illustrated in Fig.~[\ref{11}].  It can be regarded as an equal amplitude superposition of all possible $\{\sigma^z_i\}$ configurations on the A sublattice, with $\tau_q^x=-1$ at the center of cubes for which $\prod \sigma^z_i%\sigma^z_j\sigma^z_k\sigma^z_l\sigma^z_m\sigma^z_n\sigma^z_o\sigma^z_p
=-1$, 
and $\tau^x=+1$ elsewhere.
Pictured in terms of domain wall surfaces separating regions on the A sublattice with $\sigma^z = +1$ from those where $\sigma^z=-1$, this gives a $\tau_i^x=-1$ at each site on the B sublattice with an odd number of domain wall \emph{corners}.  These sites are indicated by red spins in Fig.~[\ref{11}].

We may perturb this model with transverse $\sigma^x$ and $\tau^x$ fields; when dominant these drive the system into a trivial $Z_2 \times Z_2$ paramagnet.  As for the $2D$ model, a duality transformation maps this transition  to the SSPT-breaking transition in a $3D$ version of the PIM (see Appendix \ref{app:duality}).
We may distinguish the SSPT and trivial paramagnetic phases via a \emph{volume} order parameter,
\begin{align} 
V=\langle \prod_{i \in \mathcal{C}}  \sigma^z_i \prod_{i \in \mathcal{V}} \tau^x_i   \rangle
\label{vol}
\end{align}
which is non-vanishing in the SSPT phase, but vanishes rapidly with the volume in the trivial paramagnetic phase.  
Here $\mathcal{C}$ refers to $A$ sites on the corners of a cubic volume and $\mathcal{V}$ refers to the $B$ sites in the enclosed volume. 
This nonlocal volume order parameter captures the decoration of the domain wall corners and serves as a numerical signature of the SSPT phase.

\begin{figure}[h]
  \centering
      \includegraphics[width=0.27\textwidth]{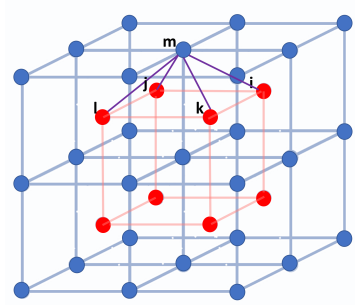}
  \caption{Geometry of the surface spin operators $\pi_i^\alpha$ for the TCIM.} 
  \label{12}
\end{figure}

For much the same reason as in $2D$, the surface of this SSPT phase has dispersionless gapless modes.  
For example, consider the surface depicted in Fig.~[\ref{12}].
We first take the Hamiltonian Eq.~[\ref{cubic}] and simply exclude terms for which the cube is not fully included in our system, as these break the subsystem symmetry.  Omitting these leaves a spin-$1/2$ degree of freedom per site on the surface, described by 
the $\pi$ Pauli matrices
\begin{align} 
\pi^x=\tau^z_i  \tau^z_j \tau^z_k \tau^z_l  \sigma^x_m,\pi^y=\tau^z_i  \tau^z_j \tau^z_k \tau^z_l  \sigma^y_m, \pi^z= \sigma^z_m
\label{edge4}
\end{align}
with $i,j,k,l,m$ as depicted in Fig.~[\ref{12}]. These surface spin operators commute with all terms in the bulk Hamiltonian, so each site on the surface has a two-fold degeneracy.   As was the case in 2D, though the subsystem symmetry operators on the system as a whole commute, their action on the degenerate Hilbert space at a single surface is effectively non-commutative.  Thus, by the same argument that applies in the 2$D$ case, 
there is no operator that can be added to the effective Hamiltonian at the surface to lift the degeneracy without breaking the $Z_2^{sub}$ symmetry.

\section{Subsystem SPT with $\mathcal{T}^{sub}$ symmetry} \label{timesub}

In the previous sections, we discussed SSPT models with a discrete $Z_2 \times Z_2$ (or more generally, $Z_m \times Z_n)$ symmetry, for which the ground state can be viewed as a ``decorated domain corner" phase, analogous to the decorated domain wall construction of global SPT phases ~\cite{chen2014symmetry}.  However this construction cannot be applied to the case of a single-component discrete symmetry.  To show that such symmetries can also lead to $d=1$ SSPTs, in this section we will present models in $D=2$ and $3$ that realizes a form of subsystem time reversal ($\mathcal{T}$) symmetry.

Time reversal is a natural symmetry choice for $d=1$ SSPTs, since it is arguably the simplest symmetry for which a $1D$ SPT phase exists\cite{Haldane1983-ya,affleck1988valence}.  
Thus an array of decoupled AKLT\cite{affleck1988valence} chains (each of which realizes the $1D$ time-reversal protected SPT\cite{Turner2011-zi}) has a subsystem symmetry in which  $\mathcal{T}$ acts on each chain individually, leading to a $\mathcal{T}$ protected Kramers doublet at the end of each chain.

However, a subtlety arises in defining anti-unitary subsystem symmetry once we weakly couple these chains -- as we must even for weak $2D$ SSPT phases.
For a spin 1/2 system, the $\mathcal{T}=\mathcal{K} i \sigma_y$ operator is a combination of the spin rotation operator $R^y=i \sigma_y$
and complex conjugation $\mathcal{K}$ (which acts on any numerical factors). 
Thus in the weakly coupled model the action of  $\mathcal{K}$ cannot be factored into a product of terms acting on separate subsystems, as the coupling introduces numerical factors that cannot be assigned to a single subsystem.  We therefore define `subsystem time reversal symmetry' ($\mathcal{T}^{sub}$) to mean symmetry under a subsystem spin rotation $R^y_{\text{sub}}= \prod_{j \in \text{sub}} \left( i \sigma^y_j \right)$ acting on all spins in the subsystem ``sub", and {\it global} complex conjugation $\mathcal{K}$.  

As we will see, this definition does allow both weak and strong SSPT phases, but with a very different type of protected boundary state than in the case of decoupled chains.  
Further, unlike in the models discussed above, here weak and strong SSPTs do not harbor different projective representations at their boundaries, but instead must be distinguished by their different bulk symmetry responses.

 \subsection{2$D$ Valence Plaquette Solid with $\mathcal{T}^{sub}$ symmetry}

\begin{figure}[h]
  \centering
      \includegraphics[width=0.26\textwidth]{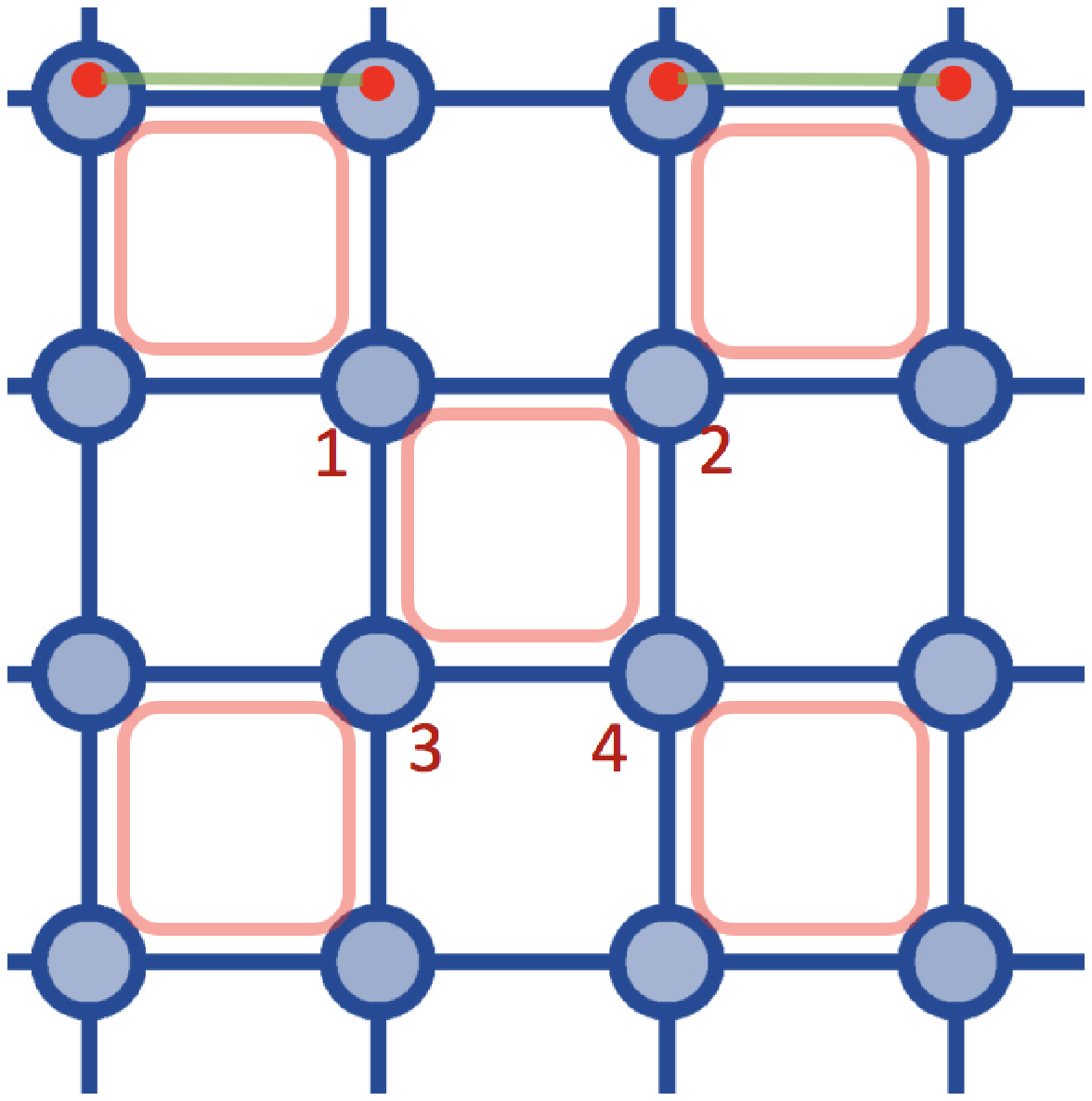}
  \caption{Our $\mathcal{T}^{sub}$ SSPT model lives on the checkboard lattice with two spins 1/2 per site.  Each red plaquette hosts a four-spin cluster interaction.  A given spin participates only in one of the two cluster interactions on neighbouring plaquettes. } 
  \label{fermion1}
\end{figure}

Our $2D$ model lives on a checkboard lattice with 2 spins-$\frac{1}{2}$ on each site, as shown in Fig.~[\ref{fermion1}]. 
Each spin interacts with one of the two red plaquette clusters $P_i$ adjacent to the site. This guarantees that, in the limit that there is no on-site interaction between the spins, the Hamiltonian is a sum of non-overlapping (and therefore commuting) plaquette clusters:
 \begin{align} 
&H=\sum_{P_i}  L_{P_i}\nonumber\\
&L_{ P}=|\alpha \rangle \langle \alpha |
\end{align}

Here $L_{P_i}$ is the plaquette cluster interaction for each individual red plaquette, which is chosen as follows.
First, it must have a unique ground state to ensure that the bulk is gapped.  Second, it must be real, and invariant under acting with $i \sigma^y$ on neighbouring pairs of spins.  Here we choose the pairs to be along the edges of the square,  
giving 
 \begin{align} 
& |\alpha \rangle=\frac{1}{\sqrt{2}}(| 0 \rangle_{12}| 1 \rangle_{34}-| 1 \rangle_{12}| 0 \rangle_{34})\nonumber\\
&=\frac{1}{\sqrt{2}}(| 0\rangle_{13}| 1 \rangle_{24}-|1  \rangle_{13}| 0 \rangle_{24})
\end{align}
where we have defined 
\begin{align}
\label{0n1}
    | 0 \rangle_{ij}\equiv \frac{1}{\sqrt{2}}[|\uparrow \rangle_i|\downarrow \rangle_j-|\downarrow \rangle_i|\uparrow \rangle_j] \nonumber \\
    | 1  \rangle_{ij}\equiv \frac{1}{\sqrt{2}}[|\uparrow \rangle_i|\uparrow \rangle_j+|\downarrow \rangle_i|\downarrow \rangle_j]
    \end{align}
Note that the expression for $|\alpha \rangle$ is the same whether we pair sites as $(1,3), (2,4)$ or as $(1,2), (3,4)$.

Since the Hamiltonian is real, $\mathcal{T}^{sub}$ symmetry acts by rotating all spins along the line by $i \sigma^y$.
 \begin{align} 
%& \mathcal{T}^{sub}=\mathcal{K} i \sigma_y, \nonumber\\
& (i \sigma^y_i) : | 0 \rangle_{ij} \rightarrow |1 \rangle_{ij},| 1  \rangle_{ij} \rightarrow -|0 \rangle_{ij}\nonumber\\
& (i \sigma^y_j) : | 0 \rangle_{ij} \rightarrow -|1 \rangle_{ij},| 1  \rangle_{ij} \rightarrow |0 \rangle_{ij}\nonumber\\
& (i \sigma^y_i)(i \sigma^y_j) : |0 \rangle_{ij} \rightarrow |0 \rangle_{ij},  |1 \rangle_{ij} \rightarrow  |1 \rangle_{ij} 
\label{Ttrans}
\end{align}
Thus $\mathcal{T}^{sub}$ along the line covering sites $1,2$ takes $|\alpha \rangle\rightarrow |\alpha \rangle$, and both the Hamiltonian and its ground state are both invariant under $\mathcal{T}^{sub}$ symmetry. 

\subsubsection{Protected gapless boundary modes}

When only plaquette projectors are included, each edge contains a spin-$1/2$ per site which is completely decoupled from the bulk.  From Eq.  (\ref{Ttrans}), we see that the spin at site $i$ behaves like a Kramers doublet under the anti-unitary $\mathcal{T}^{sub}$ symmetry, where the subsystem is a line that intersects the edge at site $i$. 
Since this line will also intersect another boundary, globally  $(\mathcal{T}^{\text{sub}})^2 = +1$ -- but on a single edge it acts projectively, via $(\mathcal{T}^{\text{sub}})^2 = -1$.

In a system of decoupled AKLT chains, where the full time-reversal symmetry may act on each chain individually, this Kramers degeneracy per site on the boundary is protected by the individual time reversal symmetries \cite{affleck1988valence}.  However in the coupled system (whether the weak SSPT phase, or the Hamiltonian given above), where the complex conjugation must be taken to be global, these symmetries are no longer independent:  if $i$ and $j$ denote subsystems that intersect the same edge at sites $i$ and $j$ respectively,
\begin{align}
    \mathcal{T}^{\text{sub}}_j = \mathcal{T}^{\text{sub}}_i U_{ij}
\end{align} 
where $U_{ij} = R^y_{\text{sub}, i} R^y_{\text{sub}, j}$ is a unitary symmetry transformation, meaning that it is a product of an {\it even} number of anti-unitary subsystem symmetries.
Thus our $\mathcal{T}^{sub}$ symmetry contains a single anti-unitary symmetry (which we many take to be $\mathcal{T}^{\text{sub}}_1$), together with a collection of unitary subsystem symmetries. Unless the unitary symmetries are realized projectively -- which is not the case for the model considered here --  $\mathcal{T}^{\text{sub}}_1$ can ensure only that a single Kramers pair is protected on each edge.  
(A more general discussion of the possible projective representations may be found in Appendix~\ref{app:groupcoho}.)

To see this, let us consider what terms can appear at the boundary of our system without breaking $\mathcal{T}^{sub}$ symmetry.  
These terms should be real, Hermitian, and invariant under conjugation by $i \sigma^y_i$ at a single site $i$ on the edge.  
Any product of an odd number of Pauli matrices is odd under global time reversal symmetry, and hence ruled out.  In addition, any product containing Pauli matrices other than $\sigma^y$ is odd under subsystem spin rotation, and hence prohibited by $\mathcal{T}^{sub}$.  Thus the operators that may be added to the edge are products of an even number of $\sigma^y_i$.

The lowest-order term that can be added is therefore:
\begin{align}
    P_{i, i+1}^{edge} = \frac{1}{2} ( 1 +  \sigma^y_i \sigma^y_{i+1} )
\end{align}
This operator is not invariant under the full action of time reversal symmetry on an individual subsystem, but is allowed in our case since complex conjugation acts globally.  
It projects the pair of spins $(i,i+1)$ into the  two-fold degenerate subspace spanned by the states $| 0 \rangle_{i,i+1} , | 1  \rangle_{i,i+1}$, giving a new effective bond- type spin 1/2 degree of freedom.  From Eq. (\ref{Ttrans}), this bond spin $1/2$  also transforms as a Kramers doublet under the anti-unitary symmetry $\mathcal{T}^{\text{sub}}$ when the subsystem is a line that intersects the edge at site $i$ or $i+1$.  

The obvious choice for our boundary Hamiltonian is therefore the classical Ising interaction $H = - \sum_i  \sigma^y_i \sigma^y_{i+1} $.  On any finite edge this retains a 2-fold ground state degeneracy; on an infinite-length boundary it spontaneously breaks the global $\mathcal{T}$ symmetry.  In neither case can it give a symmetric, gapped boundary.

Alternatively, we could begin by
adding a term $P_{i,i+1}^{edge}$ on every other bond along the edge to the Hamiltonian.  This reduces the ground state degeneracy from $2^{N_{\text{edge}}}$ to  $2^{N_{\text{edge}/2}}$, where $N_{\text{edge}}$ is the number of sites along the edge. 
We could then construct an analogous projector acting between non-overlapping pairs of the bond spins $ | 0 \rangle_{i,i+1} ,  | 1 \rangle_{i,i+1} $ to further lift the edge degeneracy to $2^{N_{\text{edge}/4}}$, and so on.  At each state a pair of Kramers doublets is combined in such a way as to leave a single Kramers doublet under $\mathcal{T}^{sub}$.  However, adding any finite number of such terms leaves a residual ground state degeneracy that grows exponentially with the boundary's length.

We emphasize that the above couplings are quite different from those allowed in phases (such as 2D fermionic topological insulators) protected by 2D global time-reversal symmetry, in which an even number of Kramers pairs on the same boundary can be gapped.  Here we find that irrespective of the initial number of Kramers pairs on the boundary, the degeneracy cannot be fully lifted.  We also find that in order to reduce the degeneracy to that of a single Kramers pair, we must create an effective spin-$1/2$ that involves all sites on the boundary, since it transforms as a Kramers doublet under $\mathcal{T}^{sub}$ acting on {\it any} line perpendicular to the boundary.

\subsubsection{Weak versus strong $\mathcal{T}^{sub}$ SSPT phases}

Since there is only a single anti-unitary symmetry for each family of subsystems, the weak and strong $\mathcal{T}^{sub}$ SSPT phases cannot harbor distinct projective representations at their boundaries (unless the unitary subsystem symmetries also act projectively, as discussed in Appendix \ref{app:groupcoho}). This is perhaps not surprising: since complex conjugation acts globally it is not clear that these projective representations are the correct quantity to characterize these boundaries.  It is also clear that the third group cohomology, which classifies $D=2$ global SPTs, also cannot be the correct  quantity, since time-reversal symmetry alone does not lead to a non-trivial bosonic SPT in $2D$\cite{Chen2011-kz}.  

However, as for the $Z_2 \times Z_2$ case, the strong $\mathcal{T}^{sub}$ SSPT introduced here is distinct from a model with the same symmetry comprised of weakly coupled AKLT chains, because the symmetry's action on the spins at each site is different: for crossed AKLT chains, the $Z_2$ component of the subsystem symmetry is spin rotation in an individual chain.  Thus spins that are part of a horizontal chain are not affected by the vertical subsystem symmetries, and vice versa.  In the strong SSPT, in contrast, all spins are flipped by one horizontal and one vertical subsystem symmetry.   

This difference is reflected in the bulk responses to certain boundary condition twists.  Suppose that we change the boundary condition for spins in every vertical subsystem from periodic to antiperiodic.  On any vertical AKLT chains, this results in an orthogonal state, since it introduces a bond that effectively carries a spin triplet, rather than a spin singlet.  For the strong SSPT however, performing a spin rotation on (say) all spins in the lower half of the system is simply the action of a subsystem symmetry.  Thus the ground state remains unchanged.  

Finally, we note that our model can equally be written (albeit not via commuting projectors) as an interaction for a single spin-$1$ degrees of freedom on each site.  A breif discussion of this construciton is given in Appendix \ref{app:Spin1}.

\subsection{3$D$ Valence Cube Solid with $\mathcal{T}^{sub}$ symmetry}

The construction of the $\mathcal{T}^{sub}$ invariant valence plaquette solid given in previous section can be generalized to higher dimensions, to yield a $d=1$ $\mathcal{T}^{sub}$ SSPT with a protected Kramers degeneracy on each boundary in any dimension.  To illustrate how the general construction works, here we present the construction for a
$\mathcal{T}^{sub}$ invariant topological paramagnet on the 3$D$ checkerboard lattice. As in the $D=2$ case, the $\mathcal{T}^{sub}$ symmetry  acts as a combination of a subsystem spin rotation  $i\sigma_y$ on a specific line in $\hat{x}, \hat{y}$, or $\hat{z}$-direction, and global complex conjugation.

\begin{figure}[h]
  \centering
      \includegraphics[width=0.45\textwidth]{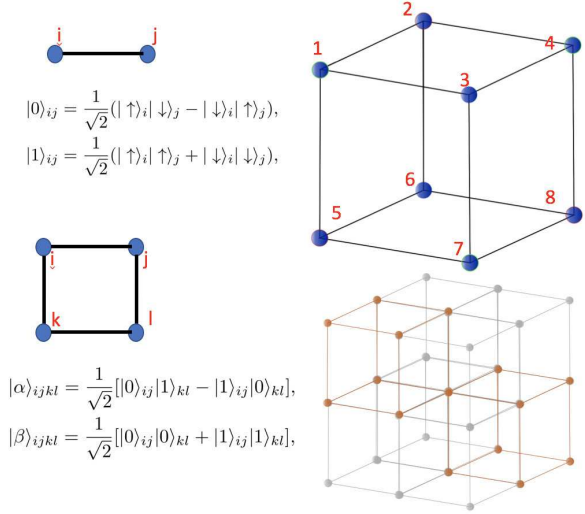}
  \caption{Left: The hierarchical construction of the ground state projector for 8 spins on a cube. First each pair of spins on a $\hat{x}$-bond is project to a 2-level subspace $|0 \rangle, |1 \rangle$.  Next a four spin interaction is introduced on the top and bottom faces of the cube, projecting each to the 2-level subspace $|\alpha \rangle, |\beta \rangle$. Finally, the 8 spins on the cube are projected to a singlet in the effective spins $|\alpha \rangle, |\beta \rangle$. Upper right: Labeling used in the main text for the 8 spins in the cluster interaction. Lower right: Each unit cell of the  3D checkerboard lattice contains eight cubes.  The two spins each participate in only one of the two cluster interactions associated with the two neighbouring red cubes.} 
  \label{xie3d}
\end{figure}

The 3D checkerboard lattice contains 8 cubes in each unit cell (Fig.~[\ref{xie3d}]). As in the $D=2$ construction, we put two spin-1/2 degrees of freedom on each site, with each spin interact with only one of the two red cubes adjacent to the site.
The Hamiltonian is a sum of  8-spin interactions on each red cube $c_i$:
 \begin{align} 
&H=\sum_{c_i}  P_{c_i}
%&L_{ijklabcd \in c}=| \chi  \rangle_{ijklabcd}  \langle \chi |_{ijklabcd} 
\label{pall}
\end{align}
Because each $P_{c_i}$  involves different spins, they are mutually commuting and the model is exactly solvable. 

The operator $P_{c_i}$ can be constructed hierarchically, by applying successive projectors to pairs of 2-state systems, as shown in Fig.~[\ref{xie3d}]. 
We begin exactly as for our $D=2$ model:
%Each site only contributes one spin 1/2 to the cluster interaction, for which we use the basis $|\uparrow \rangle,|\downarrow\rangle$.
For each cluster, we pick a pairing of spins (here we use pairs 1-2, 3-4, 5-6, 7-8, as shown in Fig.~[\ref{xie3d}]) and add to the Hamiltonian a  projector
\begin{align} 
&P_{ij}=  | 0 \rangle_{ij}\langle 0 |_{ij} + | 1 \rangle_{ij}\langle 1 |_{ij}  
\label{p1}
\end{align}
with $| 0 \rangle_{ij}, | 1 \rangle_{ij}$ defined as in Eq. (\ref{0n1}).  As in the 2D case, acting with $\mathcal{T}^{sub}$ along the line containing both sites $i$ and $j$ leaves $| 0 \rangle_{ij},| 1 \rangle_{ij}$  invariant. Acting with $\mathcal{T}^{sub}$ along the line which only crosses site $i$, the $(| 0 \rangle_{ij},|1 \rangle_{ij})$ states transform as Kramers doublet pair:
 \begin{align} 
& \mathcal{T}^{sub}=\mathcal{K} i \sigma_y, \nonumber\\
&(i \sigma^i_y) : |0 \rangle_{ij} \rightarrow |1 \rangle_{ij}, |1 \rangle_{ij} \rightarrow -|0 \rangle_{ij}
\end{align}

Next we add a projector that picks out half of the remaining states on each of two plaquettes, which we will take to be the top and bottom plaquettes of each cube (containing sites (1234) and (5678) respectively, in Fig.~\ref{xie3d}.   For the plaquette touching sites $ijkl$ %($P_{ijkl}$)
the projector is:
\begin{align} 
& P_{ijkl} = | \alpha \rangle_{ijkl} \langle  \alpha |_{ijkl} + | \beta \rangle_{ijkl} \langle  \beta |_{ijkl} 
\end{align}
with 
\begin{align}
&| \alpha \rangle_{ijkl}=\frac{1}{\sqrt{2}}[| 0 \rangle_{ij}|1\rangle_{kl}-|1 \rangle_{ij}| 0 \rangle_{kl}], \nonumber\\
& | \beta \rangle_{ijkl}=\frac{1}{\sqrt{2}}[| 0 \rangle_{ij}| 0\rangle_{kl}+|1 \rangle_{ij}|1 \rangle_{kl}]\ . 
\label{p2}
\end{align}
Once again the interaction is chosen such that acting with $\mathcal{T}^{sub}$ along a line that includes two of the sites $ijkl$ leaves the states $(| \alpha \rangle_{ijkl}, | \beta \rangle_{ijkl}) $ invariant, while under $\mathcal{T}^{sub}$ along the line perpendicular to the plaquette in question (which acts on only one of the sites$\{ ijkl \}$), $| \alpha \rangle_{ijkl}$ and $ | \beta \rangle_{ijkl}$ transform as a Kramers pair:
 \begin{align} 
& \mathcal{T}^{sub}=\mathcal{K} i \sigma_y, \nonumber\\
&(i \sigma^i_y) : |\alpha  \rangle_{ijkl} \rightarrow |\beta \rangle_{ijkl}, |\beta \rangle_{ijkl} \rightarrow -|\alpha  \rangle_{ijkl}  \ .
\end{align}
In dimension $D$, this process can be continued until an interaction between spins on a $D-1$ hypersurface is obtained.  At each step, the interaction is a projector onto a 2-fold degenerate Hilbert space.
At the last step, the projection operator picks out the unique ``singlet" ground state:
\begin{align} 
&P_{12345678}=| \chi  \rangle_{12345678}  \langle \chi |_{12345678}  , \nonumber\\
&| \chi \rangle_{12345678}=\frac{1}{\sqrt{2}}(| \alpha  \rangle_{1234}|\beta \rangle_{5678}-| \beta \rangle_{1234}|\alpha \rangle_{5678})
\label{p3}
\end{align}
The state $| \chi \rangle$ (or its analogue in higher dimensions) is invariant under $\mathcal{T}^{sub}$ symmetry,  
as now any subsystem line touches exactly two spins in any cube.  If both spins are in the same plaquette, then $|\alpha \rangle_{ijkl}$ and $|\beta \rangle_{ijkl}$ are already invariant.  If the two spins are in different plaquettes, then $|\alpha \rangle_{ijkl}$ and $|\beta \rangle_{ijkl}$ are a Kramers pair, but the singlet combination $| \chi \rangle$ is time reversal invariant.  

It is worthwhile to point out that though the construction appears to break the lattice symmetry, in fact $| \chi \rangle$ can be written equally as a singlet between top and bottom plaquettes, or between two parallel side surfaces, 
\begin{align} 
| \chi \rangle_{12345678}&=\frac{1}{\sqrt{2}}(|\alpha \rangle_{1234}|\beta \rangle_{5678}-| \beta \rangle_{1234}|\alpha \rangle_{5678}), \nonumber\\
&=\frac{1}{\sqrt{2}}(| \alpha \rangle_{1357}|\beta \rangle_{2468}-| \beta \rangle_{1357} |\alpha \rangle_{2468}), \nonumber\\
&=\frac{1}{\sqrt{2}}(| \alpha \rangle_{1256}|\beta \rangle_{3478}-| \beta \rangle_{1256}|\alpha\rangle_{3478}),
\end{align}
Hence, the eight spin state $| \chi \rangle$ can be regarded as the `all-way plaquette singlet' which projects every two parallel surfaces into the same singlet.

With $P_{c_i} = P_{12345678}$ as given in Eq. (\ref{p3}), the  Hamiltonian is clearly gapped in the bulk.  However, as in our $2D$ model, 
each  plaquette on the system's boundary that belongs to a red cube contains 4 dangling spins, and hence a Kramers pair along any subsystem that ends on this surface.  As for $D=2$, since there is only one independent anti-unitary symmetry, $\mathcal{T}^{sub}$ protects only a single Kramers degeneracy across the entire surface.   For example, we may define projectors into the subspace $|\alpha\rangle_{ijkl} ,|\beta \rangle_{ijkl}$ for each surface plaquette, reducing the surface degeneracy to a two-level system per plaquette, which transforms as a Kramers doublet under $\mathcal{T}^{sub}$ on any line perpendicular to the surface.

\section{A Fermionic SSPT}\label{fermionsub}
\label{fsspt}

Thus far, we have explored the zoology of subsystem SPT phases in interacting bosonic (spin) systems.  We have shown that for on-site unitary symmetries, these phases realize physics very similar to that of crossed arrays of decoupled 1d SPT's, while for anti-unitary time reversal symmetry, which is not strictly on-site, the symmetry-protected boundary degeneracy is much reduced, though nonetheless distinct from the global symmetry case.  We now turn to the question of whether, and how, these ideas apply to fermionic subsystem SPT phases.  

Because our main tool is to study exactly solvable model Hamiltonians, fermions introduce a new technical challenge: a fermion parity subsystem symmetry requires interactions that are at least quartic in the fermion operators.  
The resulting Hamiltonians are generally not solvable unless the interaction terms treat non-overlapping sets of fermions --  and necessarily not in the same symmetry class as any non-interacting topological phases of fermions. Thus in the fermionic context we will be more limited in our ability to construct models whose physics can be easily understood, as we have done in bosonic systems.
Here we will give one example, building a 2D Majorana model with subsystem fermion parity symmetry and global $\mathcal{T}$ symmetry, which we show is a fermion parity protected SSPT.  Our Hamiltonian contains only four-body Majorana interactions, and the resulting ground state can be interpreted as charge $4e$ superconductivity, with order parameter $\Delta_{4e} = \langle \psi^{\dagger}_k\psi^{\dagger}_{k'}\psi_{-k} \psi_{-k'} \rangle$.

\begin{figure}[h]
  \centering
      \includegraphics[width=0.26\textwidth]{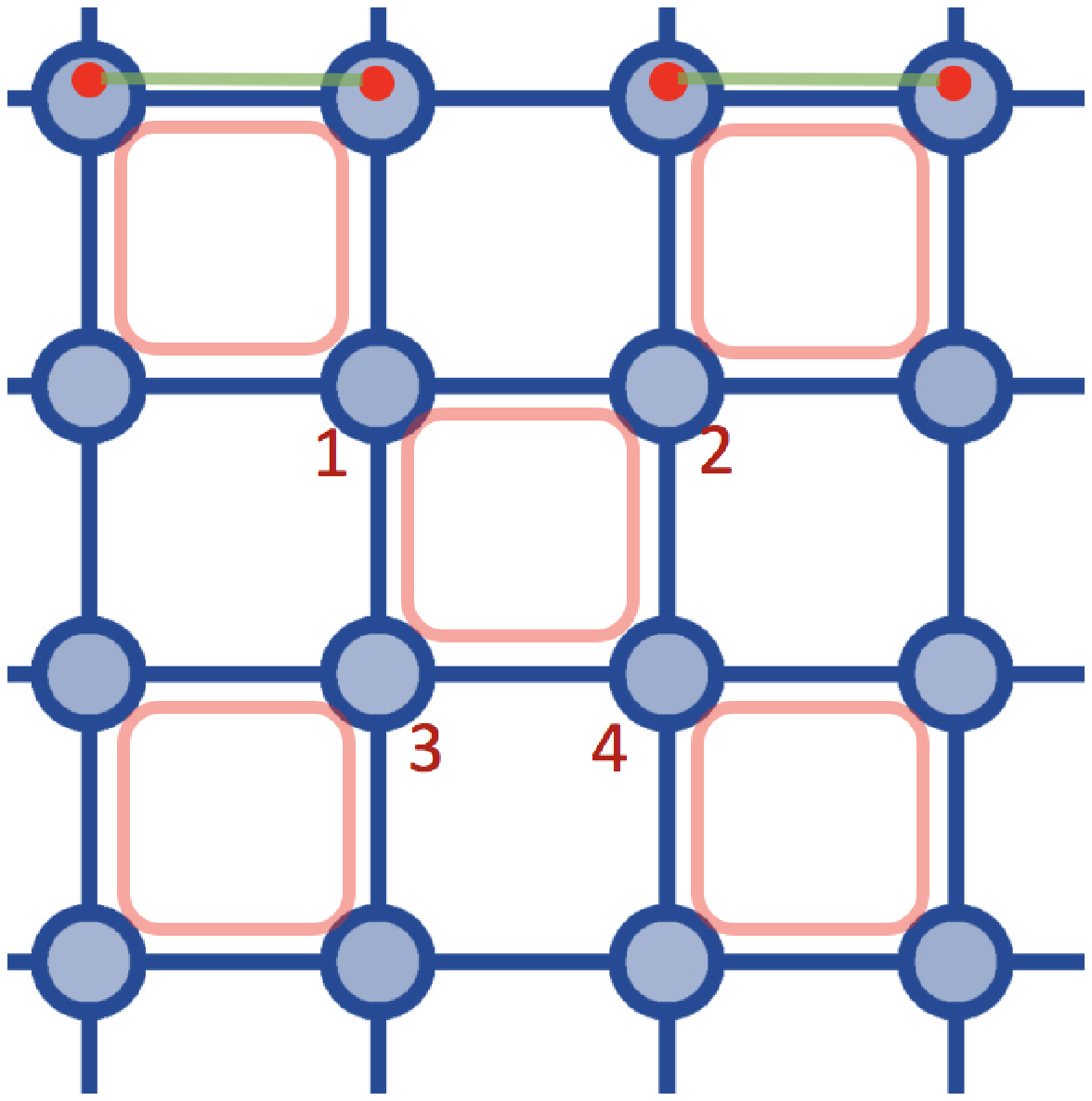}
  \caption{Our fermionic SSPT lives on the checkboard lattice, with two complex fermions per site.  Red squares indicate plaquettes hosting 4-fermion cluster interactions.  Each fermion participates in the  cluster interaction of only one of the two neighbouring plaquettes.} 
  \label{fermion}
\end{figure}
We begin with 2 fermions on each site of the checkerboard lattice.  On each checkerboard square (red squares in Fig.~[\ref{fermion}]) there is a four-body interaction between the fermions at the plaquette corners. Of the two fermions on each site, each participates in the interaction term for only one red plaquette, so that all interactions commute.

 To describe the interaction terms, we label the four fermions involved in the interaction for plaquette $P$ as $\psi_1,\psi_2,\psi_3,\psi_4$ (see Fig.~[\ref{fermion}]) and decompose each fermion into two Majoranas via $\psi_i=\eta_i+i\chi_i$. 
 Each plaquette cluster thus contains 8 Majorana fermions, which we couple via Fidkowski-Kitaev\cite{fidkowski2010effects,fidkowski2011topological} type interactions.
 Specifically, 
we first add a 4-Majorana interaction
\begin{align} 
H_1=\chi_1\chi_2\chi_3\chi_4+\eta_1\eta_2\eta_3\eta_4 \ .
\label{four}
\end{align}
Ground states of $H_1$ can be described via the bond fermions
\begin{align} 
&\Psi_{\uparrow}=\chi_1+i\chi_2,\Psi_{\downarrow}=\chi_3+i\chi_4\nonumber\\
&\Psi'_{\uparrow}=\eta_1+i\eta_2,\Psi'_{\downarrow}=\eta_3+i\eta_4
\label{mapf}
\end{align}
In these variables, the Hamiltonian $H_1$ becomes,
\begin{align} 
H_1=(n_{\Psi}-1)^2+(n_{\Psi'}-1)^2
\end{align}
Thus $H_1$ favors the odd fermion parity state for both $\Psi$ and $\Psi'$. 
This allow us to map the ground state subspace of $H_1$ into two spin $1/2$ degrees of freedom per plaquette:
\begin{align} 
&\vec{n}_i=\Psi^{\dagger} \vec{\sigma}_i \Psi, \vec{m}_i=\Psi'^{\dagger} \vec{\sigma}_i \Psi', 
\end{align}

In terms of these spin degrees of freedom, the second interaction on the plaquette cluster is
\begin{align} 
& H_2=- \vec{m} \cdot \vec{n}\nonumber\\
&= (\chi_1\chi_2-\chi_3\chi_4)(\eta_1\eta_2-\eta_3\eta_4)\nonumber\\
&+(\chi_1\chi_4-\chi_2\chi_3)(\eta_1\eta_4-\eta_2\eta_3)\nonumber\\
&+(\chi_1\chi_3+\chi_2\chi_4)(\eta_1\eta_3+\eta_2\eta_4)
\end{align}
This anti-ferromagnetic interaction   
projects the two spins in each plaquette cluster into a singlet, yielding a unique ground state. With this cluster interaction on each red plaquette in Fig.~[\ref{fermion}], the many-body Hamiltonian is fully gapped with a unique ground state in the bulk.

What are the symmetries of this model?  First, it has an anti-unitary symmetry $\mathcal{T} = \mathcal{K}$, which acts by global complex conjugation, taking 
\begin{align} \label{FermionT}
  \mathcal{T}:  \chi \rightarrow \chi, \eta \rightarrow \eta, i \rightarrow -i \ \ .
\end{align}  On the physical fermions $\psi_i$ at each site (or the bond fermions $\Psi_{i, \sigma}$) this gives a particle-hole transformation; it takes the spin vectors $\vec{m}_i \rightarrow - \vec{m}_i, \vec{n}_i \rightarrow - \vec{n}_i$ (see Appendix \ref{fermiapp} for details). 
Thus on both sets of operators, $\mathcal{T}^2 =+ 1$.  However, the transformation of the operators $\vec{m}_i, \vec{n}_i$ (which play the role of the Pauli matrices for our spin-$1/2$) implies that a spin-$1/2$ state transforms projectively, with $\mathcal{T}^2 = -1.$
The plaquette interaction, and its resulting spin-singlet ground state, are clearly $\mathcal{T}$ invariant.

In addition, our Hamiltonian conserves the fermion parity of each row/column separately.  The corresponding fermion parity symmetry operator is   
\begin{align} 
Z^{\text{fp}, \text{sub}}_2=&e^{i\pi \sum_{i }n_{\psi_i}} \nonumber \\
Z^{fp,sub}_2:& \chi,\eta \rightarrow -\chi,-\eta %\nonumber\\
%&\mathcal{T},\chi,\eta \rightarrow \chi,\eta,i  \rightarrow -i
\end{align}
which clearly leaves $H_1$ invariant.  Acting on the vertical line crossing sites $1,3$ this subsystem $Z^{fp,sub}_2$ symmetry acts on the spins via:
\begin{align} 
%&Z^{fp,sub}_2:\chi_1,\eta_3 \rightarrow -\chi_1,-\eta_3 \nonumber\\
&Z^{fp,sub}_2: (n^x_1,n^y_2,n^z_3) \rightarrow (-n^x_1,-n^y_2,n^z_3), \nonumber\\
&(m^x_1,m^y_2,m^z_3) \rightarrow (-m^x_1,-m^y_2, m^z_3)
\end{align}
Likewise, $Z^{fp,sub}_2$ symmetry acting on the horizontal line crossing sites $1,2$ takes
\begin{align} \label{FpHorizontal}
&Z^{fp,sub}_2: (n^x_1,n^y_2,n^z_3) \rightarrow (n^x_1,-n^y_2,-n^z_3), \nonumber\\
&(m^x_1,m^y_2,m^z_3) \rightarrow (m^x_1,-m^y_2,-n^z_3)
\end{align}
The anti-ferromagnetic interaction between the two effective spins in the plaquette cluster thus respects the  $Z^{\text{fp,sub}}_2$ symmetry.

With only the plaquette interaction, in the presence of an edge each boundary site contains a free complex fermion (or two free Majoranas).  In 1 dimension a boundary mode of this type is protected by time-reversal symmetry \cite{fidkowski2010effects},   which prohibits quadratic interactions between Majoranas (or more generally, interactions between $4n +2$ Majoranas, which must have an imaginary pre-factor in order to be hermitian).  

Since we cannot gap the system without breaking $\mathcal{T}$ by coupling the pair of Majoranas at a single site on the boundary, we must consider what can be done by coupling multiple sites at the boundary.  Four body interactions of the form 
\begin{align}
\label{First4f}
    \chi_{i-1} \chi_i \eta_{i-1}\eta_{i}
\end{align}
are allowed by symmetry.  Since these operators square to 1, each individual term lifts the 4-fold degeneracy of a quartet of free Majoranas to a 2-fold degeneracy, which can be viewed as a spin-$1/2$ degree of freedom.

Unlike for the bulk, however, we cannot gap these boundary  spin-$1/2$ degrees of freedom by coupling pairs of them into singlets, as one can for a single $1D$ chain\cite{fidkowski2010effects,levin2012classification}. This is because the remaining 4-fermion interactions required to couple the spin-$1/2$'s to form singlets in the ground state violate $Z^{\text{fp,sub}}_2$.

Indeed, as discussed following Eq. (\ref{FermionT}), the effective spin-$1/2$ per unit cell that is left after introducing the interaction (\ref{First4f})  transforms as a Kramers doublet under time reversal. (For details, see Appendix \ref{fermiapp}.)  Here time reversal is a global symmetry, so this alone does not guarantee a gapless boundary.  Instead, consider a product of time reversal symmetry and a subsystem fermion parity operation along all subsystems that intersect the boundary {\it except one}:
\begin{align}
    \tilde{T} = \mathcal{T} \prod_{j \neq i}Z^{fp, j}
\end{align}
Since $\mathcal{T}$ anticommutes with the fermion parity symmetry when acting on a single row that intersects the boundary, this operator can be viewed as a product of subsystem symmetries that square to $1$ with a single anti-unitary subsystem symmetry that squares to $-1$.  
Thus we find ourselves in essentially the same situation as in the previous section, with a single symmetry-protected Kramers degeneracy on each boundary.

The previous discussion suggests a close connection between this fermion SSPT and our previous boson SSPT models with $\mathcal{T}^{sub}$.
Though some phases of matter are uniquely fermionic, some phases of interacting fermions can also be realized in bosonic systems.  In this sense the fermionic model discussed here realizes a phase very similar to the $\mathcal{T}^{sub}$ bosonic SSPT, albeit with an extra $Z_2$ spin-rotation subsystem symmetry.    
The nature of the boundary degeneracy, and of the action of the symmetries both on the bulk and at the boundary, is the same in both models. 
(For details, see Appendix \ref{fermiapp}.)
This is not entirely surprising, since the interactions effectively couple plaquette fermions in such a way that the low-energy Hilbert space can be described by two spins$-1/2$ per plaquette. 

This leaves open the interesting question of how to realize an SSPT which is fundamentally fermionic, in the sense that it does not have any bosonic equivalent.

\section{Concluding remarks}
In this work we propose a new type of symmetry protected topological matter: SSPT phases, whose gapless boundary modes are protected by subsystem symmetries. We have established the existence of this class of phases by constructing explicit examples in three classes - two bosonic and one fermionic. In doing so we have expanded the understanding of phase structure in the presence of subsystem symmetries which was previously restricted to broken symmetry phases. We remind the reader that subsystem symmetries have come to the fore recently in the study of fracton phases which exhibit subdimensional particle motion and emergent higher-rank gauge fields \cite{pretko2017fracton,Pretko2017-ej,Pretko2017-nt}.

Our work raises three immediate interesting questions.  First, we have focused exclusively on  SSPT phases protected by $d=1$ subsystem symmetries. In three dimensions, one may also consider $d=2$ subsystem symmetries, which are of particular interest due to their close connection to fracton topological order\cite{Vijay2016-dr}.  In a companion paper we will explore $d=2$ SSPT phases in $3D$. Second, we have not systematically analyzed the dependence of SSPT phases on lattice structure in a given physical and subsystem dimension. Third, what is the classification of SSPT phases?  We have shown  that for some unitary symmetries, the SSPT phase is associated with projective representations of the boundary spins that are distinct from what can be realized by any array of $D=1$ global SPTs.  As shown in Appendix \ref{app:groupcoho}, this cannot happen for familiar continuous symmetries such as SO(3), U(1), and SU(2). However, it is not clear whether this is a {\it defining} characteristic of any SSPT, or merely an interesting feature of our models.  For antiunitary and fermionic symmetries the classification is even less clear, since the full symmetry is effectively a mixture of unitary subsystem symmetry and global complex conjugation, which appears to be naturally classified neither by projective representations (i.e. $\mathcal{H}^2$) or invariants associated with $2D$ global SPTs. We look forward to progress on these inter-related issues.

\begin{acknowledgments}
We are grateful to S. Parameswaran, R. Nandkishore, A. Prem and R.Roy for insightful comments and discussions. Y-Y is supported by a PCTS Fellowship at Princeton University.  FJB is grateful for the financial support of NSF-DMR 1352271 and the Sloan Foundation FG-2015-65927.  SLS is supported by DOE Grant No. DE-SC/0016244.

\end{acknowledgments}

\appendix

\section{Group cohomology calculation of strong SSPT boundaries} \label{app:groupcoho}

As discussed in the main text, one situation in which there are certainly distinct weak and strong SSPT phases is when the boundary of the strong SSPT  realizes the symmetry in a way that cannot be obtained from decoupled (or weakly coupled) 1D SPT's. Here we give group cohomological calculations for some familiar symmetry groups to illustrate when such representations exist.  The upshot of our calculation is that they do exist for symmetry groups, such as $Z_2$ or more generally $Z_n$, where certain torsions associated to their group cohomology are non-vanishing.  They do not exist for many familiar symmetry groups, such as U(1) and SO(3).

Consider a unitary symmetry group $G$, which can be either a finite group or a compact Lie group.  The number of projective representations is given by\cite{Chen2013-gq}:
\be
\mathcal{H}^2(G, U(1)) = \mathcal{H}^3(G, \mathbb{Z} ) 
\ee
Thus what we need to evaluate in general is the third group cohomology with integer coefficients, $\mathcal{H}^3(G, \mathbb{Z} ) $.  
To do this we use the Kunneth formula:
\begin{align}
 &\mathcal{H}^d (G_1 \times G_2 , \mathbb{Z} )= \nonumber\\
 & \prod_{p=0}^d\left [ \mathcal{H}^p ( G_1,  \mathbb{Z} ) \otimes_{\mathbb{Z}}\mathcal{H}^{d-p}  ( G_2,  \mathbb{Z} ) \right ]  \n
% \left [ \mathcal{H}^1 ( G_1,  \mathbb{Z} ) \otimes_{\mathbb{Z}}\mathcal{H}^2 ( G_1,  \mathbb{Z} ) \right ]  \left [ \mathcal{H}^2 ( G_1,  \mathbb{Z} ) \otimes_{\mathbb{Z}}\mathcal{H}^1 ( G_1,  \mathbb{Z} ) \right ]  \left [ \mathcal{H}^3 ( G_1,  \mathbb{Z} ) \otimes_{\mathbb{Z}}\mathcal{H}^0 ( G_1,  \mathbb{Z} ) \right ] 
& \prod_{p=0}^{d+1 }\left [  \text{Tor}_1^{  \mathbb{Z}} ( \mathcal{H}^p ( G_1,  \mathbb{Z} ),\mathcal{H}^{d+1-p}  ( G_2,  \mathbb{Z} ) \right ]
\end{align}
Detailed information about the meaning of this notation can be found in Ref. \onlinecite{Chen2013-gq}; for our purposes the relevant facts are that
for any $M$,
\be
 \mathbb{Z} \otimes_{\mathbb{Z}} M = M \otimes_{\mathbb{Z}}  \mathbb{Z}  = M \ , \ \ \  \mathbb{Z}_1 \otimes_{\mathbb{Z}} M = M  \otimes_{\mathbb{Z}} \mathbb{Z}_1  =  \mathbb{Z}_1
 \ee
 and
 \be
 \text{Tor}_1^{  \mathbb{Z}} ( \mathbb{Z}, M ) =  \text{Tor}_1^{  \mathbb{Z}} ( \mathbb{Z}_1, M )  =   \mathbb{Z}_1 
 \ee
 with $\text{Tor}_1^{  \mathbb{Z}} (A,  B) = \text{Tor}_1^{  \mathbb{Z}} (B, A)$.  

In the examples that we tabulate here, we will begin from groups for which
\be
 \mathcal{H}^0 ( G,  \mathbb{Z} ) =  \mathbb{Z} \ , \ \  \mathcal{H}^1 ( G,  \mathbb{Z} ) =  \mathbb{Z}_1 
 \ee
It follows that for any $n$ 
\begin{align}
 \mathcal{H}^0 ( G^n,  \mathbb{Z} ) =  \mathbb{Z} \ , \ \  \mathcal{H}^1 ( G^n,  \mathbb{Z} ) =  \mathbb{Z}_1 \nonumber \\
  \mathcal{H}^2 ( G^n,  \mathbb{Z} ) = \left(   \mathcal{H}^2 ( G,  \mathbb{Z} )  \right)^n
 \end{align}
 This is easily seen using the Kunneth formula: we have
\begin{align}
&\mathcal{H}^0 (G_1 \times G_2, \mathbb{Z} ) =\nonumber\\
&  \mathcal{H}^0 ( G_1,  \mathbb{Z} ) \otimes_{\mathbb{Z}}\mathcal{H}^{0}  ( G_2,  \mathbb{Z} )  \n
&  \prod_{p=0}^{1 }\left [  \text{Tor}_1^{  \mathbb{Z}} ( \mathcal{H}^p ( G_1,  \mathbb{Z} ),\mathcal{H}^{1-p}  ( G_2,  \mathbb{Z} ) \right ]
\end{align}
Thus if $ \mathcal{H}^0 ( G_1,  \mathbb{Z} ) =  \mathcal{H}^0 ( G_2,  \mathbb{Z} ) = \mathbb{Z}$, then $ \mathcal{H}^0 ( G_1 \times G_2,  \mathbb{Z} )  =  \mathbb{Z}$.  Thus if $ \mathcal{H}^0 ( G,  \mathbb{Z} ) = \mathbb{Z}$, then $ \mathcal{H}^0 ( G^n,  \mathbb{Z} ) = \mathbb{Z}$.
Further, 
\begin{align}
&\mathcal{H}^1 (G_1 \times G_2 , \mathbb{Z} )=\nonumber\\
& \prod_{p=0}^1\left [ \mathcal{H}^p ( G_1,  \mathbb{Z} ) \otimes_{\mathbb{Z}}\mathcal{H}^{1-p}  ( G_2,  \mathbb{Z} ) \right ]  \n
&  \prod_{p=0}^{2 }\left [  \text{Tor}_1^{  \mathbb{Z}} ( \mathcal{H}^p ( G_1,  \mathbb{Z} ),\mathcal{H}^{d+1-p}  ( G_2,  \mathbb{Z} ) \right ]
\end{align}
 which is clearly the trivial group $\mathbb{Z}_1$ if $ \mathcal{H}^1 ( G_1,  \mathbb{Z} ) = \mathcal{H}^1 ( G_2,  \mathbb{Z} ) =  \mathbb{Z}_1$.  Thus if $ \mathcal{H}^1 ( G,  \mathbb{Z} ) = \mathbb{Z}_1$, then also $ \mathcal{H}^1 ( G^n,  \mathbb{Z} ) = \mathbb{Z}_1$.
 
 Under these assumptions, we have:
 \begin{align}
\mathcal{H}^2 (G_1 \times G_2 , \mathbb{Z} ) &= &  \mathcal{H}^2 ( G_1,  \mathbb{Z} ) \times \mathcal{H}^{2}  ( G_2,  \mathbb{Z} )   
\end{align}
 since all relevant torsions involve either an $\mathcal{H}^0$ or $\mathcal{H}^1$, and therefore vanish.  
 
Next, we use the Kunneth formula to evaluate $\mathcal{H}^3 (G_1 \times G_2 , \mathbb{Z} )$, where $\mathcal{H}^1 (G_1, \mathbb{Z} ) = \mathcal{H}^1 (G_2, \mathbb{Z} ) = \mathbb{Z}_1$, and $\mathcal{H}^0 (G_1, \mathbb{Z} ) = \mathcal{H}^0 (G_2, \mathbb{Z} ) = \mathbb{Z}$.  In this case, the Kunneth formula reduces to:
 \begin{align}
\mathcal{H}^3 (G_1 \times G_2 , \mathbb{Z} ) &= & \mathcal{H}^{3}  ( G_2,  \mathbb{Z} ) \times \mathcal{H}^3 ( G_1,  \mathbb{Z} ) \n
& & \left [  \text{Tor}_1^{  \mathbb{Z}} ( \mathcal{H}^2 ( G_1,  \mathbb{Z} ),\mathcal{H}^{2}  ( G_2,  \mathbb{Z} ) )\right ]
\end{align}

To understand this formula, let us first consider the case that $\text{Tor}_1^{  \mathbb{Z}} ( \mathcal{H}^2 ( G,  \mathbb{Z} ),\mathcal{H}^{2}  ( G,  \mathbb{Z} ) ) = \mathbb{Z}_1$.  We must now apply a second fact about torsion:
\be
\text{Tor}_1^{  \mathbb{Z}} (A \times B, M ) = \text{Tor}_1^{  \mathbb{Z}} (A, M) \times \text{Tor}_1^{  \mathbb{Z}} (B, M)
\ee
Thus if $A = \mathcal{H}^2 ( G,  \mathbb{Z} )$, and $\text{Tor}_1^{  \mathbb{Z}} ( A, A) =  \mathbb{Z}_1$, then
\be
\text{Tor}_1^{  \mathbb{Z}} ( A\times A, A ) = \mathbb{Z}_1
\ee
By induction it follows that $\text{Tor}_1^{  \mathbb{Z}} ( A^n, A ) = \mathbb{Z}_1$, so that the torsion term is always trivial.  In this case 
\begin{align}
\mathcal{H}^3 (G^n , \mathbb{Z} ) &= &  \left [ \mathcal{H}^{3}  ( G,  \mathbb{Z} )\right ]^n \ \ \ .
\end{align}
This means that any projective representation on a boundary that intersects $n$ subsystems can be obtained by taking $n$ decoupled $1d$ $G$-SPTs, and the only possibility is a weak SPT phase.
This is the case, for example, for $G=$SU(2), SO(3), or U(1), .

To understand what happens when the torsion is non-vanishing, we will consider the concrete example of $G = \mathbb{Z}_p$.  In this case 
\begin{equation}
 \mathcal{H}^{2}  ( \mathbb{Z}_p,  \mathbb{Z} )    =  \mathbb{Z}_p
\end{equation}
 and we can use (yet another) fact about torsion:
 \begin{equation}
 \text{Tor}_1^{  \mathbb{Z}} (  \mathbb{Z}_p,  \mathbb{Z}_q ) = \mathbb{Z}_{(p,q)}
\end{equation}
where $(p,q)$ is the greatest common divisor of $p$ and $q$.  It follows that
\begin{align}
\text{Tor}_1^{  \mathbb{Z}} (  \mathbb{Z}_p \times   \mathbb{Z}_p ,  \mathbb{Z}_p ) = (\mathbb{Z}_{p})^{2} \ , \\ 
\text{Tor}_1^{  \mathbb{Z}} (   \mathbb{Z}_p \times ( \mathbb{Z}_p )^2 ,  \mathbb{Z}_p   ) = (\mathbb{Z}_{p})^{3}  \nonumber \\
\text{Tor}_1^{  \mathbb{Z}} (   \mathbb{Z}_p \times ( \mathbb{Z}_p )^{n-1} ,  \mathbb{Z}_p   ) = (\mathbb{Z}_{p})^{n} 
\end{align}
where the last line follows by induction.  

Then we have:
\begin{equation}
\mathcal{H}^3 ( \mathbb{Z}_p  ^{n-1} \times  \mathbb{Z}_p   , \mathbb{Z} ) = \mathcal{H}^3 (  \mathbb{Z}_p ^{n-1},  \mathbb{Z} ) 
\times ( \mathbb{Z}_{p})^{n-1}
\end{equation}
from which we find that;
 \begin{align}
  \mathcal{H}^{3}  (  \mathbb{Z}_p ,  \mathbb{Z} ) = \mathbb{Z}_1 \n 
    \mathcal{H}^{3}  (  \mathbb{Z}_p^2  ,  \mathbb{Z} ) =  \mathbb{Z}_p \n
    \mathcal{H}^{3}  (  \mathbb{Z}_p^3  ,  \mathbb{Z} ) = ( \mathbb{Z}_p)^3 \n
     \mathcal{H}^{3}  (  \mathbb{Z}_p^4  ,  \mathbb{Z} ) = ( \mathbb{Z}_p)^6 \n
      \mathcal{H}^{3}  (  \mathbb{Z}_p^n  ,  \mathbb{Z} ) = ( \mathbb{Z}_p)^{\frac{n^2-n}{2}}
\end{align}
 In particular, for every $n$, there are projective representations which cannot be equated to taking copies of $q=n/m$ projective representations of  $(\mathbb{Z}_p)^{m}$ (for any $m$), since $ \frac{1}{2} ( ( m q)^2 - mq ) > \frac{q}{2} (m^2 -m) $.    These must be projective representations in which the boundary spins do not transform as independent clusters under the symmetries.  It is in this case that SSPT phases may exist.  
 
 To establish the existence of these phases we must understand in more detail the projective representation in question, and how it acts on all spins in an interdependent manner.  We do not undertake to enumerate the possibilities here, except to note that the construction in the main text gives one example.

\subsection{Projective representations of $G^n \times \mathbb{Z}_2^T$}

The other class of example that we discuss in our paper involves, effectively, two types of symmetries: global complex conjugation and a unitary symmetry $G$.  (In the examples of the main text, $G=\mathbb{Z}_2$.)  The group cohomology of pure time reversal symmetry (realized as complex conjugation, with no spin rotation) is
\begin{align}
\mathcal{H}^{0}  ( \mathbb{Z}_2^T , \mathbb{Z}^T )= \mathcal{H}^{2}  ( \mathbb{Z}_2^T , \mathbb{Z}^T ) = \mathbb{Z}_1  \nonumber \\
\mathcal{H}^{1}  ( \mathbb{Z}_2^T , \mathbb{Z}^T )= \mathcal{H}^{3}  ( \mathbb{Z}_2^T , \mathbb{Z}^T ) = \mathbb{Z}_2
\end{align}
In this case, one can use the Kunneth formula and our previous calculation of $\mathcal{H}^d(G^n, \mathbb{Z})$ to obtain the result in (what should be) a relatively straightforward manner.
\begin{align}
&\mathcal{H}^d (G^n \times \mathbb{Z}_2^T,\mathbb{Z}^T)\nonumber\\
& =  \prod_{p=1,3,...}^d\left [ \mathcal{H}^{d-p} ( G^n,  \mathbb{Z} ) \otimes_{\mathbb{Z}}\mathcal{H}^{p}  ( \mathbb{Z}_2^T , \mathbb{Z}^T ) \right ]  \n
& \prod_{p=1,3,... }^{d+1 }\left [  \text{Tor}_1^{  \mathbb{Z}} ( \mathcal{H}^{d+1-p} ( G^n,  \mathbb{Z} ),\mathcal{H}^{p}  ( \mathbb{Z}_2^T , \mathbb{Z}^T ) \right ]
\end{align}
where the contributions involving $\mathcal{H}^{2p} ( \mathbb{Z}_2^T , \mathbb{Z}^T ) =  \mathbb{Z}_1$ are trivial.  
Thus 
\begin{align}
\mathcal{H}^0 (G^n \times \mathbb{Z}_2^T , \mathbb{Z}^T) &= &
\left [  \text{Tor}_1^{  \mathbb{Z}} ( \mathcal{H}^{0} ( G^n,  \mathbb{Z} ),\mathcal{H}^{1}  ( \mathbb{Z}_2^T , \mathbb{Z}^T ) \right ]
\nonumber \\
&= &
  \text{Tor}_1^{  \mathbb{Z}} (   \mathbb{Z} ,  \mathbb{Z}_2)  = \mathbb{Z}_1
\end{align}

\begin{align}
\mathcal{H}^1 (G^n \times \mathbb{Z}_2^T , \mathbb{Z}^T) &= & \left [ \mathcal{H}^{0} ( G^n,  \mathbb{Z} ) \otimes_{\mathbb{Z}}\mathcal{H}^{1}  ( \mathbb{Z}_2^T , \mathbb{Z}^T ) \right ]  \n
& & \left [  \text{Tor}_1^{  \mathbb{Z}} ( \mathcal{H}^{1} ( G^n,  \mathbb{Z} ),\mathcal{H}^{1}  ( \mathbb{Z}_2^T , \mathbb{Z}^T ) \right ]  \n
& = &  \mathbb{Z}_2 
\end{align}
where as above we have assumed that $\mathcal{H}^0 (G,  \mathbb{Z}) =  \mathbb{Z}, \mathcal{H}^1 (G,  \mathbb{Z}) =  \mathbb{Z}_1$.

Finally,
 \begin{align}
\mathcal{H}^3 (G^n \times \mathbb{Z}_2^T , \mathbb{Z}^T) &= & \left [ \mathcal{H}^{2} ( G^n,  \mathbb{Z} ) \otimes_{\mathbb{Z}} \mathbb{Z}_2 \right ]  \times \mathbb{Z}_2  \n
& & \left [  \text{Tor}_1^{  \mathbb{Z}} ( \mathcal{H}^{3} ( G^n,  \mathbb{Z} ), \mathbb{Z}_2 ) \right ] \n
& & \left [  \text{Tor}_1^{  \mathbb{Z}} ( \mathcal{H}^{1} ( G^n,  \mathbb{Z} ), \mathbb{Z}_2 ) \right ] \n
&=& \left [ (  \mathcal{H}^{2} ( G,  \mathbb{Z} ) )^n  \otimes_{\mathbb{Z}} \mathbb{Z}_2 \right ]  \times ( \mathbb{Z}_2) \n
& & \left [  \text{Tor}_1^{  \mathbb{Z}} ( \mathcal{H}^{3} ( G^n,  \mathbb{Z} ), \mathbb{Z}_2 ) \right ]
\end{align}
Here the factor of $\mathbb{Z}_2$ arises from the assumption that $\mathcal{H}^{1} ( G,  \mathbb{Z} ) = \mathbb{Z} $.  The remaining factors depend on the specifics.

For $G = \mathbb{Z}_2$ (the example discussed in the text), $\mathcal{H}^{2} ( G,  \mathbb{Z} ) = \mathbb{Z}_2$, while $\mathcal{H}^{3} ( G^n,  \mathbb{Z} ) = \mathbb{Z}_2^{\frac{n^2 -n}{2} }$.  This gives a total of $\mathbb{Z}_2^{\frac{n^2 +n +2}{2} }$ possible projective representations.
Because the difference between this number and the number allowed for $Z_2^n$ alone is a power of $n$, we conclude that any new projective representations associated with time-reversal symmetry (rather than with $Z_2$) can be realized as boundary states of weak SSPTs.  This is consistent with our finding that the strong and weak $T$-SSPTs cannot be distinguished by the symmetry action on their boundaries.

\section{Duality between SPT and symmetry breaking state} \label{app:duality}

An interesting feature of many models with Ising-like symmetry is the existence of duality mappings, such as Kramers-Wannier duality~\cite{Kramers1941-po}, relating different parameter regimes or models. Such duality mappings are useful in understanding the structure of each model's phase diagram, and in understanding its phase transitions.

In this section, we outline a duality that maps our SSPT models perturbed with a transverse field to two copies of the trivial plaquette Ising model. Since the latter is known to be self-dual, this also implies a self-duality transformation for our SSPT.
Like with the Kramers-Wannier duality for the Ising model~\cite{Kramers1941-po}, this self-duality allows us to pinpoint the critical point exactly.

\subsection{Duality between 2D SSPT and plaquette Ising model}

According to our previous construction, the subsystem SPT phase protected by 1$d$ symmetry can always be reached by coupled SPT chains. Suggestively, the 1$d$ $ZXZ$\cite{chen2014symmetry} chain can be mapped to a transverse Ising model via a nonlocal duality transformation\cite{Xu2004-oj}.  Here, we show that our 2$d$(3$d$) subsystem SPT with transverse field is also dual to the plaquette (cube) Ising model with a transverse field.  Under this duality the subsystem SPT phase is mapped to the subsystem symmetry breaking phase, and the edge modes in subsystem SPT become the global degeneracy associated with spontaneous symmetry breaking. 

For convenience, we repeat the Hamiltonian for our topological plaquette Ising paramgnet protected by $Z^{sub}_2\times Z^{sub}_2$ in Eq.~[\ref{topo1}] with a transverse field:
\begin{align} 
&H=-\sum_{ijklm \in P_a} \sigma^z_i \sigma^z_j \sigma^z_k \sigma^z_l  \tau^x_m-\sum_{ijklm \in P_b} \tau^z_i \tau^z_j \tau^z_k \tau^z_l  \sigma^x_m
\nonumber\\
&+\Gamma\tau^x+\Gamma\sigma^x
\label{dual1}
\end{align}
For small $\Gamma$, the system is in the topological plaquette paramagnetic phase; at large $\Gamma$ the spins polarize and drive the system into the trivial paramagnetic phase.
We now define a nonlocal transformation to map the model into a plaquette Ising model with transverse field,
\begin{figure}[h]
  \centering
      \includegraphics[width=0.24\textwidth]{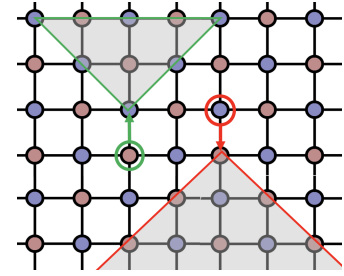}
  \caption{The non-local duality transformation dresses each $\sigma^z$($\tau^z$) operator with a cone of $\tau^x$($\sigma^x$) indicated by the shaded region inside the red(green) triangle.} 
  \label{dual2}
\end{figure}

\begin{align} 
&\sigma^x  \rightarrow \sigma^x,\tau_x  \rightarrow \tau^x;\nonumber\\
&\sigma^z  \rightarrow \sigma^z \prod_{i \in red} \tau_i^x, \tau^z  \rightarrow \tau^z \prod_{i \in green} \sigma^x_i
\end{align}
The duality transformation dresses each $\sigma^z$($\tau^z$) operator with a cone of $\tau^x$($\sigma^x$) operators inside the shaded region in the red(green) triangle. After the transformation, the Hamiltonian becomes,
\begin{align} 
&H=-\sum_{ijklm \in P_a} \sigma^z_i \sigma^z_j \sigma^z_k \sigma^z_l  -\sum_{ijklm \in P_b} \tau^z_i \tau^z_j \tau^z_k \tau^z_l  
\nonumber\\
&+\Gamma\tau^x+\Gamma\sigma^x
\label{eq:dual2}
\end{align}
which is exactly a pair of plaquette Ising models in transverse fields. The SSPT state in our original model becomes the subsystem symmetry breaking phase after duality. 
The nonlocal membrane order parameter in Eq.~[\ref{mem}] characterizing the decorated corner nature of the SSPT becomes the four point correlator of plaquette Ising model which measures the subsystem symmetry breaking order,
\begin{align} 
&O=\langle \prod_{ijkl \in \mathcal{C}}  \sigma^z_i \sigma^z_j  \sigma^z_k \sigma^z_l \prod_{abc \in \mathcal{M}} \tau^x_a \tau^x_b \tau^x_c ...  \rangle \nonumber\\
&\rightarrow O=\langle \sigma^z_i \sigma^z_j  \sigma^z_k \sigma^z_l \rangle
\end{align}

The degeneracy of the subsystem-symmetry broken phase is represented, in the dual SSPT model, by the degeneracy of the protected gapless boundary modes.  To see this, consider gapping one of these boundary modes by adding a term $\sigma_z\tau_x \sigma_z$ to the Hamiltonian at the boundary, as shown in Fig.~[\ref{three}].   Duality maps this term to $\sigma_z \sigma_z$ on an edge, which breaks the subsystem symmetry and reduces the ground state degeneracy. After some simple counting, we can conclude that the $4^{L_x}4^{L_y-1}$ edge modes becomes the $4^{L_x}4^{L_y-1}$ fold degeneracy of the plaquette Ising model (two copies) with subsystem symmetry breaking.

As the plaquette Ising model with transverse field is self-dual, with a single first-order transition  at $\Gamma=1$\cite{Xu2004-oj,Xu2005-df,Orus2009-zh,Kalis2012-io},%This model has two distinct phases with a  at $\Gamma=1$~\cite{}. 
it follows that the SSPT phase and trivial phase in Eq.~[\ref{dual1}] is also self-dual with a topological transition happening at $\Gamma=1$.  Indeed a self-duality transformation for our model can be constructed as follows.
We define the controlled-Z (CZ) operator acting on a two-spin Hilbert space in the $z$-basis to be the diagonal matrix $\text{CZ} = \text{diag}(1,1,1,-1)$.
Letting  $U$ denote the unitary obtained by performing CZ on all the bonds, one can show that
\begin{eqnarray}
U H(\Gamma) U^\dagger = h H(\Gamma^{-1})
\end{eqnarray}
where 
\begin{equation}
    H(\Gamma) = H_{\text{TPIM(TCIM)}} - \Gamma \sum_{i\in A} \sigma_i^x - \Gamma \sum_{i\in B}\tau^x_i
\end{equation}
for the 2D (TPIM) or 3D (TCIM) models.

\subsection{Duality between 3D SSPT and cube Ising model} 
\begin{figure}[h]
  \centering
      \includegraphics[width=0.4\textwidth]{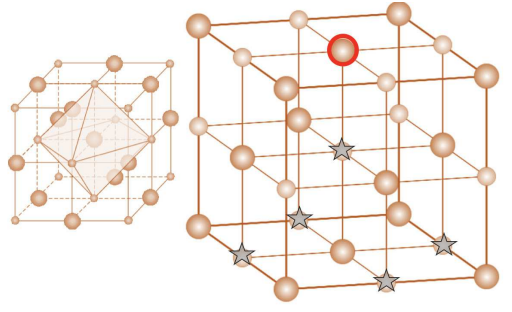}
  \caption{L:The 3D cubic paramagnet on the Ion lattice with two type of sites. The unit cell of each type forms an octahedron. The large site $\tau$ stays in the center of the octahedron while the small site $\sigma$ lives on the corner of octahedron. R: During the non-local duality, each $\sigma_z$(red circle) operator is dressed with a set of $\tau_x$ operators inside the pyramid below (grey starred sites). Meanwhile, each $\tau_z$ operator is dressed with a set of $\sigma_x$ operators in the pyramid above.} 
  \label{dual}
\end{figure}

A similar approach yields a duality between 3D topological cubic paramagnet and a subsystem-symmetry broken phase in a trivial $Z^{sub}_2\times Z^{sub}_2$-symmetric model. To simplify the mapping, we consider the Hamiltonian for the cubic paramagnetic phase on the Ion-lattice with two type of sites on a simple cubic lattice, as in Fig.~[\ref{dual}]. The large/ small sites refer to two type of Pauli spins $\tau,\sigma$. The unit cell forms an octahedron with 6 corner sites.
On this lattice, the Hamiltonian of the topological cubic paramagnet with a transverse field is,
\begin{align} 
H=&-\sum_{ijklmnq \in O_A}  \sigma^z_i \sigma^z_j \sigma^z_k \sigma^z_l\sigma^z_m \sigma^z_n  \tau^x_q+\Gamma\tau^x \nonumber\\
& -\sum_{ijklmnq \in O_B} \tau^z_i \tau^z_j \tau^z_k \tau^z_l \tau^z_m \tau^z_n \sigma^x_q+\Gamma\sigma^x 
\end{align}
Here $O_A,O_B$ are the octahedra for $A$ and $B$ sublattices, respectively. The cluster interaction contains six $\sigma_z$ spins on the corners of the 
octahedron and one $\tau_x$ in the center of octahedron. Such a cluster interaction decorates the domain wall corner with transverse spin $\tau_x$. When $\Gamma$ is small, the system is in the topological cubic paramagnetic phase. 

The duality mapping is:
\begin{align} 
&\sigma_x  \rightarrow \sigma_x,\tau_x  \rightarrow \tau_x;\nonumber\\
&\sigma_z  \rightarrow \sigma_z \prod_{i \in P_a} \tau^i_x, \tau_z  \rightarrow \tau_z \prod_{i \in P_b} \sigma^i_x
\end{align}
This transformation dress each $\sigma_z$($\tau_z$) operator with a set of $\tau_x$($\sigma_x$) operators inside the Pyramid region, as shown in Fig.~[\ref{dual}]. After the transformation, the Hamiltonian becomes the cubic Ising model with a transverse field.
\begin{align} 
H=&-\sum_{ijklmn \in O_a}  \sigma^z_i \sigma^z_j \sigma^z_k \sigma^z_l\sigma^z_m \sigma^z_n  +\Gamma\tau^x \nonumber\\
& -\sum_{ijklmn \in O_b} \tau^z_i \tau^z_j \tau^z_k \tau^z_l \tau^z_m \tau^z_n+\Gamma\sigma^x 
\end{align}

As for the $2D$ model, the duality maps the subsystem SPT ground state to the subsystem symmetry breaking ground state. The protected edge modes and volume order become the ground state degeneracy and six point order parameter which characterizes the symmetry breaking, respectively.

\section{Spin 1 $\mathcal{T}^{sub}$ SSPT on checkerboard lattice} \label{app:Spin1}

In section \ref{timesub}, we construct an exactly solvable model for $\mathcal{T}^{sub}$ SSPT on checkerboard lattice with two independent spin 1/2 per site. If we project the two onsite spins into $S=1$ subspace, the system is still in the same SSPT state with $\mathcal{T}^{sub}$ symmetry although the Hamiltonian is no longer solvable.

After enforcing $S=1$ per site, we introduce the Schwinger boson representation for the spin system,
\begin{align} 
&a^{\dagger}_ia_i+b^{\dagger}_ib_i=2,a^{\dagger}_ia_i-b^{\dagger}_ib_i=2S_i^z,\nonumber\\
&S^+_i=a^{\dagger}_ib_i,S^-_i=b^{\dagger}_ia_i
\end{align}

The $|\alpha \rangle$ state projection in each plaquette becomes,
\begin{align} 
&|\alpha \rangle=[(a^{\dagger}_1b^{\dagger}_2-b^{\dagger}_1a^{\dagger}_2)(a^{\dagger}_3a^{\dagger}_4+b^{\dagger}_3b^{\dagger}_4)\nonumber\\
&-(a^{\dagger}_3b^{\dagger}_4-b^{\dagger}_3a^{\dagger}_4)(a^{\dagger}_1a^{\dagger}_2+b^{\dagger}_1b^{\dagger}_2)]
|0 \rangle
\end{align}

The ground state wave function is the product of each plaquette cluster entangled as $|\alpha \rangle$. The edge state contains an effective spin 1/2 each site which cannot be fully gapped without breaking $\mathcal{T}^{sub}$.

\section{Majorana fermion cluster interaction} \label{fermiapp}
In this appendix, we provide details of the Majorana fermion cluster interaction in the fermion SSPT model in Section.~[\ref{fsspt}].

Using the definitions  
\begin{align} 
&\Psi_{\uparrow}=\chi_1+i\chi_2,\Psi_{\downarrow}=\chi_3+i\chi_4\nonumber\\
&\Psi'_{\uparrow}=\eta_1+i\eta_2,\Psi'_{\downarrow}=\eta_3+i\eta_4\nonumber\\
&n_i=\Psi^{\dagger} \sigma_i \Psi, m_i=\Psi'^{\dagger} \sigma_i \Psi'
\end{align}
from the main text, the explicit expressions for 
 the O(3) rotor degrees of freedom $n,m$ is:
\begin{align} 
&n_1=-i\chi_1\chi_2+i\chi_3\chi_4,\nonumber\\
&n_2=-i\chi_1\chi_4+i\chi_2\chi_3,\nonumber\\
&n_3=i\chi_1\chi_3+i\chi_2\chi_4,\nonumber\\
&m_1=-i\eta_1\eta_2+i\eta_3\eta_4,\nonumber\\
&m_2=-i\eta_1\eta_4+i\eta_2\eta_3,\nonumber\\
&m_3=i\eta_1\eta_3+i\eta_2\eta_4,
\end{align}
From these expressions it is straightforward to derive the action of symmetry operations on these spin degrees of freedom.  First, clearly global $\mathcal{T}$ symmetry acts on the rotor via:
\begin{align} 
\mathcal{T}: \vec{n},\vec{m} \rightarrow  -\vec{n},-\vec{m} \ .
\end{align}
The subsystem $Z^{fp,sub}_2$ symmetry acting on a vertical line crossing sites $1,3$ takes $Z^{fp,sub}_2:\chi_1,\eta_3 \rightarrow -\chi_1,-\eta_3$, whence we derive:
\begin{align} 
&Z^{fp,sub}_2: (n_1,n_2,n_3) \rightarrow (-n_1,-n_2,n_3), \nonumber\\
&(m_1,m_2,m_3) \rightarrow (-m_1,-m_2,n_3)
\end{align}
as claimed in the main text.
A similar argument shows that $Z^{fp,sub}_2$ symmetry acting on a horizontal line crossing sites $1,2$ acts via Eq. (\ref{FpHorizontal}).

Let us now turn to the action of our symmetries at the system's edge.  
In the fermion model, before adding boundary terms each site on the edge contains a single fermion zero mode $\psi=\eta+i\chi$. The interaction $\eta_{2i} \chi_{2i} \eta_{2i+1} \chi_{2i+1}$ is allowed by symmetry, and reduces the fermion zero modes on two edge sites to a spin 1/2 degree of freedom:
\begin{align} 
&n'^z_{i}=-i\chi_{2i}\eta_{2i}+i\chi_{2i+1}\eta_{2i+1},\nonumber\\
&n'^y_{i}=-i\chi_{2i}\eta_{2i+1}+i\chi_{2i+1}\eta_{2i},\nonumber\\
&n'^x_{i}=i\chi_{2i}\chi_{2i+1}+i\eta_{2i+1}\eta_{2i}
\end{align}
The global $\mathcal{T}$ symmetry acts on the rotor as,
\begin{align} 
\mathcal{T}: \vec{n} \rightarrow  -\vec{n}
\end{align}
The subsystem $Z^{fp,sub}_2$ symmetry, when acting on the vertical line crossing the edge becomes an onsite Fermion-parity symmetry. Such onsite Fermion-parity symmetry becomes the onsite $Z_2$ symmetry acting on the site spin as,
\begin{align} 
&Z^{fp,sub}_2: (n^x_{i},n^y_{i},n^z_{i}) \rightarrow (-n^x_{i},-n^y_{i},n^z_{i})
\end{align}
Such $Z^{fp,sub}_2$ symmetry prohibits any local edge spin interaction in the $S_x-S_y$ plane. The only allowed interaction, without breaking $\mathcal{T}$ and $Z^{fp,sub}_2$ symmetry is the classical Ising spin interaction $\sigma_z(i)\sigma_z(i+1)...$ whose ground state would break the $\mathcal{T}$ symmetry.

\end{document}